# Large Itinerant Electron Exchange Coupling in the Magnetic Topological Insulator MnBi$_2$Te$_4$

Hari Padmanabhan,[1] Vladimir A. Stoica,[1] Peter Kim,[2] Maxwell Poore,[2] Tiannan Yang,[1] Xiaozhe Shen,[3] Alexander H. Reid,[3] Ming-Fu Lin,[3] Suji Park,[3] Jie Yang,[3] Huaiyu Wang,[1] Nathan Z. Koocher,[4] Danilo Puggioni,[4] Lujin Min,[1] Seng-Huat Lee,[5,6] Zhiqiang Mao,[5,6] James M. Rondinelli,[4] Aaron M. Lindenberg,[3,7] Long-Qing Chen,[1] Xijie Wang,[3] Richard D. Averitt,[2] John W. Freeland,[8] Venkatraman Gopalan[1]

[1]Department of Materials Science and Engineering, The Pennsylvania State University, University Park, PA 16802, USA
[2]Department of Physics, University of California San Diego, La Jolla, CA 92093, USA
[3]SLAC National Accelerator Laboratory, Menlo Park, CA 94025, USA
[4]Department of Materials Science and Engineering, Northwestern University, Evanston, IL 60208, USA
[5]2D Crystal Consortium, Materials Research Institute, The Pennsylvania State University, University Park, PA 16802, USA
[6]Department of Physics, Penn State University, University Park, PA 16802, USA
[7]Department of Materials Science and Engineering, Stanford University, Menlo Park, CA 94305, USA
[8]Argonne National Laboratory, Lemont, IL 60439, USA

**Magnetism in topological materials creates phases exhibiting quantized transport phenomena with applications in spintronics and quantum information. The emergence of such phases relies on strong interaction between localized spins and itinerant states comprising the topological bands, and the subsequent formation of an exchange gap. However, this interaction has never been measured in any intrinsic magnetic topological material. Using a multimodal approach, this exchange interaction is measured in MnBi$_2$Te$_4$, the first realized intrinsic magnetic topological insulator. Interrogating nonequilibrium spin dynamics, itinerant bands are found to exhibit a strong exchange coupling to localized Mn spins. Momentum-resolved ultrafast electron scattering and magneto-optic measurements reveal that itinerant spins disorder via electron-phonon scattering at picosecond timescales. Localized Mn spins, probed by resonant X-ray scattering, disorder concurrently with itinerant spins, despite being energetically decoupled from the initial excitation. Modeling the results using atomistic simulations, the exchange coupling between localized and itinerant spins is estimated to be >100 times larger than superexchange interactions. This implies an exchange gap of >25 meV should occur in the topological surface states. By directly quantifying local-itinerant exchange coupling, this work validates the materials-by-design strategy of utilizing localized magnetic order to create and manipulate magnetic topological phases, from static to ultrafast timescales.**

## 1. Introduction

Magnetic order in functional quantum materials provides a rich platform for the discovery of cooperative phenomena with applications in spintronics, magnetic memory, quantum information science, and beyond. Such phenomena arise out of interaction between magnetic order and other relevant microscopic degrees of freedom. This is exemplified in the Mn(Bi,Sb)$_{2n}$Te$_{3n+1}$ family of materials, which are the first realization of intrinsic magnetic order in a topological insulator.[1] The band topology is found to be intimately connected to magnetism, with experimental demonstration of quantum anomalous Hall and axion insulator phases,[2,3] and magnetic field-driven topological phase transitions.[3,4] While the magnetic order is due to localized Mn $3d$ spins, it is the exchange coupling between the localized spins and the topological surface states,[5] and the resultant formation of an exchange gap in the surface states that is responsible for the above phenomena.

The topological surface states in Mn(Bi,Sb)$_{2n}$Te$_{3n+1}$ are the product of band inversion in bulk, $p$-like Bi/Sb and Te states.[6] The exchange gap $m$ induced in these surface states by magnetic order can be written in the form $m = J_{pd}\bar{S}_z$ (see **Figure 1a**) where $\bar{S}_z$ is the projection of Mn $3d$ localized spins along the normal to the surface and $J_{pd}$ is the exchange coupling between localized Mn $3d$ spins and the $p$-like carriers that comprise the surface states.[5,7–9] The bulk and surface magnetic properties have been comprehensively characterized using a variety of experimental techniques including magnetometry,[10–

[12] neutron scattering,[10,12,13] and magnetic force microscopy.[14,15] On the other hand, despite its outsized influence on magnetic topological phenomena, experimental insight into $J_{pd}$ remains elusive. Magnon dispersion measured by inelastic neutron scattering may allow experimental access to such interatomic exchange couplings, however, the absence of a discernible *p*-like magnetic moment makes such an approach infeasible.[10] Magneto-transport measurements have been used to estimate such interactions in dilute magnetic semiconductors through the effects of carrier-spin scattering.[16,17] However, this is an indirect probe that relies on simplifying assumptions on exchange interactions which do not hold in intrinsic magnetic materials. Such indirect approaches may furthermore be obscured by other interactions such as magnetodielectric coupling.[18] As a consequence, no experimental quantification of $J_{pd}$ and its influence on magnetic dynamics has been reported on the Mn(Bi,Sb)$_{2n}$Te$_{3n+1}$ family of materials, or indeed any intrinsic magnetic topological material.

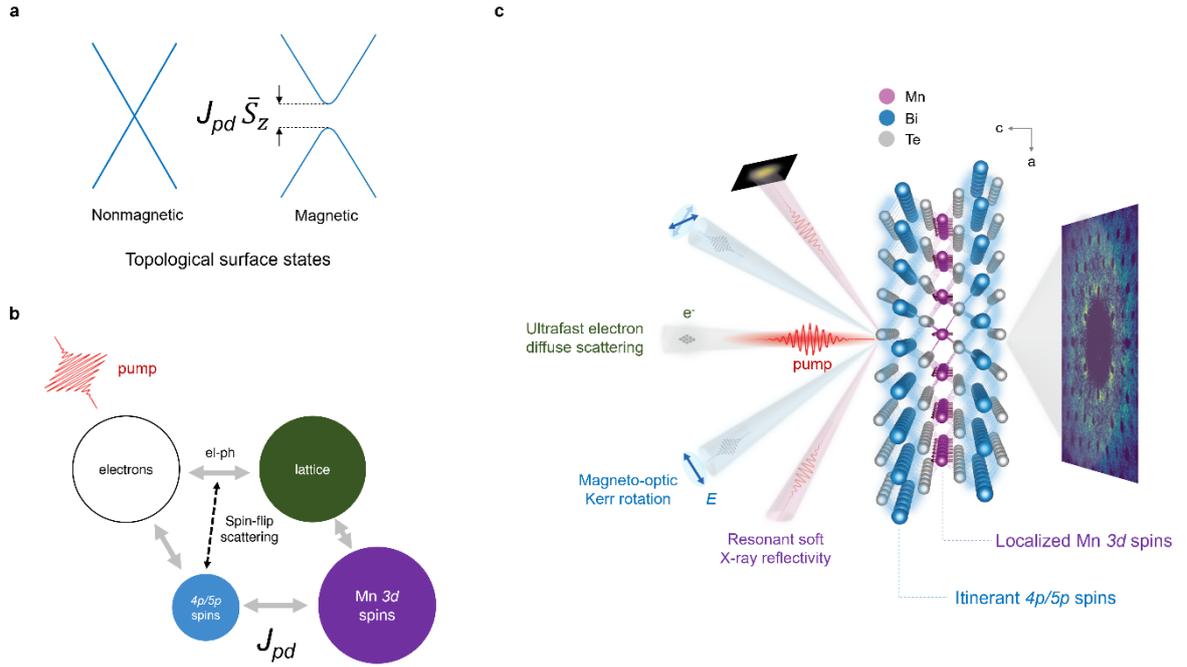

**Figure 1 Multimodal probe of nonequilibrium magnetic dynamics. a,** Schematic of the topological surface states in a magnetic topological material, with the exchange gap given by $J_{pd}\bar{S}_z$, where $J_{pd}$ is the *p-d* exchange coupling and $\bar{S}_z$ is the projection of Mn *3d* localized spins along the normal to the surface. **b,** Schematic illustrating the interplay between nonequilibrium electronic, spin, and lattice dynamics in MnBi$_2$Te$_4$, with $J_{pd}$ highlighted. **c,** Schematic of the topological surface states in a magnetic topological material, with the exchange gap given by $J_{pd}S$. **c,** Schematic illustrating the time-resolved experimental probes used to characterize ultrafast spin and lattice dynamics in MnBi$_2$Te$_4$. The crystal lattice structure of MnBi$_2$Te$_4$ is shown with the itinerant *p* spins illustrated schematically using blue clouds, and localized *d* spins illustrated using arrows on Mn ions. Ultrafast electron diffuse scattering is used to probe electron-phonon scattering and nonequilibrium phonon dynamics. Magneto-optic Kerr rotation (with the polarization of the optical electric field *E* illustrated using blue arrows) is used to probe dynamics of itinerant Bi and Te *p* spins. Resonant soft X-ray reflectivity at the Mn *L* edge is used to probe dynamics of localized Mn *d* spins.

The quantification of $J_{pd}$ is especially critical in light of the unresolved debate over the existence of an exchange gap in the topological surface states in Mn(Bi,Sb)$_{2n}$Te$_{3n+1}$ and related materials. While transport experiments clearly reveal signatures of quantized behavior due to magnetism,[2,3] angle-resolved photoemission spectroscopy (ARPES) measurements show that the exchange gap in the surface states is zero within the experimental resolution,[19–21] despite the well-established bulk magnetic order. Some of the explanations proposed to explain this discrepancy are possible reconstruction of magnetic order at the surface,[19] nanometer scale magnetic domains,[20] and vanishingly weak $J_{pd}$ exchange coupling.[14] Recent surface magnetic measurements contradict the former proposals,[14,15,21] whereas the latter has no supporting evidence. The necessity of directly experimentally evaluating $J_{pd}$ thus becomes urgent.

We interrogate $J_{pd}$ in the intrinsic antiferromagnetic (AFM) topological insulator MnBi$_2$Te$_4$ by driving the system out of equilibrium and measuring the magnetic response at its intrinsic timescales. Ultrafast optical excitation is used to create hot carriers, which thermalize with the lattice and spins through various channels (see Figure 1b) and subsequently melt the magnetic order. We employ a multimodal ultrafast approach (see Figure 1c) to characterize the nonequilibrium spin dynamics and their interplay with electrons and the lattice (shown schematically in Figure 1b). Ultrafast electron diffuse scattering in combination with magneto-optic Kerr effect measurements reveal clear signatures of spin-polarized Bi and Te $p$-like itinerant bands, which disorder through electron-phonon spin-flip scattering processes at picosecond timescales. Localized Mn $3d$ spins, probed through resonant soft X-ray scattering, despite being located far below the Fermi level and energetically decoupled from the initial optical excitation are observed to disorder concurrently with the itinerant spins. On the other hand, in the absence of hot carriers which demagnetize the itinerant spins, the localized spins instead melt at orders-of-magnitude slower rates via thermalization with the lattice. The above observations suggest that the Mn $3d$ localized spins and bulk $p$-like itinerant bands are strongly coupled via a strong interatomic exchange coupling $J_{pd}$. Modeling our experimental results using orbital-resolved atomistic Landau-Lifshitz-Gilbert (LLG) simulations, we estimate a $J_{pd}$ that is ~100 times larger than the in-plane superexchange coupling and predict that the topological surface states should exhibit an exchange gap >25 meV. Our multimodal measurement of ultrafast magnetic dynamics in MnBi$_2$Te$_4$ thus sheds light on the exchange pathways that enable magnetic topological phases and provides a foundation for the exploration of spin-based phenomena interfaced with band topology.

## 2. Results

### 2.1. Momentum-resolved probe of electron-phonon scattering

A key ingredient of any description of nonequilibrium magnetic dynamics is the transfer of angular momentum associated with ordered spins into different degrees of freedom. In an ultrafast demagnetization experiment, charge carriers scattering with phonons is hypothesized to be the dominant channel of dissipation of spin angular momentum.[22–24] Here, every electron-phonon scattering event is associated with a finite spin-flip probability, where the hot electron provides the energy to flip the spin and the phonon provides the angular momentum. Given this, a complete description of ultrafast magnetic dynamics requires the quantification of electron-phonon coupling and nonequilibrium phonon dynamics. We employ ultrafast electron diffuse scattering (UEDS) as a direct probe of electron-phonon and phonon-phonon scattering. UEDS provides a momentum- and mode-resolved measure of phonon populations, overcoming limitations of purely optical approaches. As we show below, the phonon scattering rate can vary by as much as a factor of 5 across the Brillouin zone, highlighting the importance of a momentum-resolved technique. Our results show that hot electrons in MnBi$_2$Te$_4$ are directly coupled to optical phonons, which subsequently couple to acoustic phonons. In particular, we show that carrier-optical phonon scattering results in demagnetization via spin-flip processes.

A snapshot of the ultrafast electron diffuse scattering, obtained by subtracting the static diffuse scattering intensity from the time-integrated intensity at time delays $t > 0$ is shown in **Figure 2a** (also see Methods and SI Section S1). Phonon Brillouin zones (BZs) are overlayed on the scattering phase space as a guide to the eye. The diffuse scattering intensity $I(\boldsymbol{Q},\boldsymbol{q}) \propto \frac{n_j(\boldsymbol{q})+\frac{1}{2}}{\omega_j(\boldsymbol{q})}|F(\boldsymbol{Q})|^2$, where $\boldsymbol{Q}$ is the scattering wavevector, $\boldsymbol{q}$ is the reduced scattering wavevector within the BZs marked in Figure 2a, $j$ is the phonon branch index, $n$ is the phonon population, and $F$ is the phonon structure factor.[25] A full mode-resolution of the transient scattering intensity is hindered by the large number of phonons (3 acoustic and 18 optical branches) and their entwined dispersion curves.[26] A more informative approach is to separate the acoustic and optical phonon contributions. Acoustic phonons, due to their vanishing frequencies near the zone center, will dominate the UEDS signal (see SI Section S2). In Figure 2b, we show a simulation of the structure factor $F_{TA}(\boldsymbol{Q})$ of the transverse acoustic (TA) phonon branch, which exhibits a divergent intensity near the zone-centers with a characteristic azimuthal dependence. This azimuthal form factor follows from the dependence of $F_j(Q)$ on the dot product $Q \cdot e_{j,s}(q)$ where $e_{j,s}(q)$ is the eigendisplacement of ion $s$ in phonon branch $j$ at reduced wavevector $q$ (see SI Section S2). Examination of the experimental data in Figure 2a shows a clear signature of this mode. It is noteworthy that there are no signatures of the longitudinal acoustic branch, which has a radial form factor (see SI Section S2 for simulation and discussion). Finally, the diffuse scattering intensity in regions where $F_{TA}(Q)$ vanishes may be attributed to optical phonons (see SI Section S2 for a detailed discussion of phonons away from the zone-center). Based on this, we employ regions of integration as shown in the schematic in Figure 2c to isolate the contribution of optical (green) and acoustic (grey) phonons to the diffuse scattering intensity.

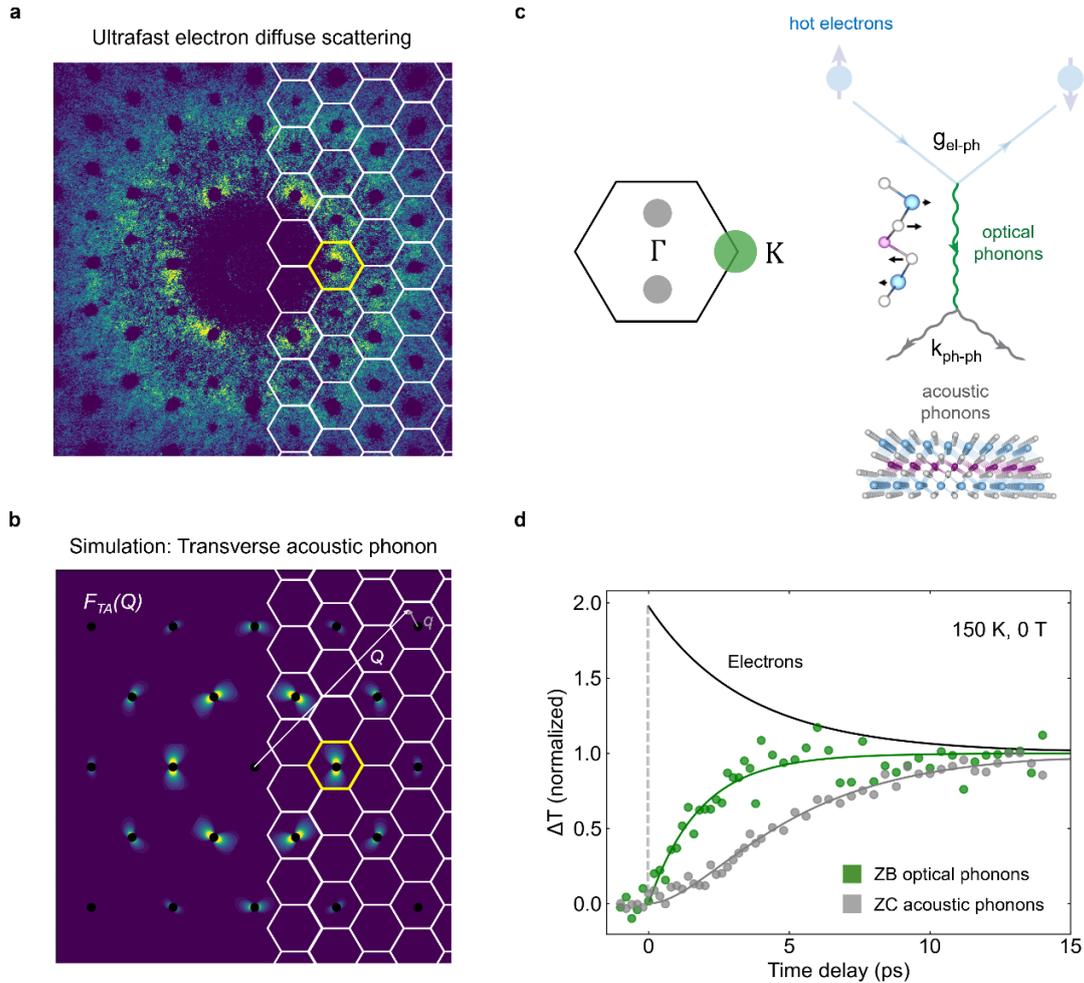

**Figure 2 Momentum-resolved electron-phonon scattering. a,** Snapshot of ultrafast electron diffuse scattering, obtained by subtracting the static diffuse scattering from the pump-induced diffuse scattering integrated over time delay $t > 0$. **b,** Simulation of the structure factor $F_{TA}(Q)$ of the transverse acoustic phonon branch, where $Q$ is the scattering wavevector. Brillouin zones (BZ) are drawn in white as a guide to the eye in (a) and (b). **c,** (right) Schematic of electron-phonon and phonon-phonon scattering, with the respective coupling constants $g_{el-ph}$ and $k_{ph-ph}$. (left) Regions of integration to obtain optical (green) and acoustic (grey) phonon populations for the BZ highlighted in yellow in (a) and (b). **d,** Experimental phonon populations (circles). These were obtained using the regions of integration shown in (c). The solid lines are fits to the non-thermal model shown schematically in panel (c).

We carry out measurements at 150 K and above where transient phonon population changes may be approximated as temperature changes, i. e. $\Delta n \propto \Delta T$ (see SI Section S3). While this is well above the AFM ordering temperature $T_N = 24$ K, we note that electron-phonon coupling is very weakly temperature-dependent in the range of electronic temperatures achieved in our experiments, and in the absence of electronic and structural phase transitions.[27–29] Indeed, as we show below, despite the difference in temperatures, our UEDS measurements are fully consistent with low-temperature ultrafast magnetic measurements.

The plot in Figure 2d shows the experimental transient (ZB = zone boundary) optical phonon and (ZC = zone center) acoustic populations converted to normalized temperatures (see SI Section S3) as a function of time delay after pump excitation. The results show a dramatic difference in the pump-induced transient phonon populations. In particular, in the first two picoseconds, only optical phonons show an appreciable transient population, with acoustic phonons exhibiting a delayed rise. We analyze the results using a nonthermal model of nonequilibrium phonons, consisting of two phonon subsystems (optical and acoustic) individually described by distinct transient temperatures. The pump excitation is assumed to instantaneously create an electronic state with an elevated temperature, which subsequently thermalizes with the lattice by scattering with the two phonon subsystems. The scattering of hot electrons with optical phonons is described by the electron-phonon coupling parameter $g_{el-ph}$. The optical phonons decay into acoustic phonons by a phonon-phonon coupling parameter $k_{ph-ph}$, as shown in the schematic in Figure 2c. We fit the entire set of temperature-dependent experimental data (see SI Section S3 for other temperatures) to a temperature-independent $g_{el-ph}$ and a variable $k_{ph-ph}$. The details of our numerical simulations and fitting can be found in SI Section S3. For the experiments at 150 K, the initial electronic temperature upon pump excitation is approximately 1340 K, and the final temperature upon equilibration is around 750 K. The fit results, plotted as solid lines in Figure 2d, closely follow the experimental results, and correspond to $g_{el-ph} = 0.27 \times 10^6$ J K$^{-1}$ m$^{-3}$ ps$^{-1}$ and $k_{ph-ph} = 0.1 \times 10^6$ J K$^{-1}$. Our final fit results are robust, remaining unchanged with one order-of-magnitude variation in the initial fit estimates.

The following observations are made – (i) the experimental data is consistent with a model in which hot electrons couple only to optical phonons, i. e. the fits of the numerical simulations are completely insensitive to any additional coupling between hot electrons and acoustic phonons. (ii) The thermalization of the phonon subsystem takes over 5 ps, as opposed to the electron-optical phonon scattering timescale of around 1 ps. (iii) The experimental data is consistent with a temperature-independent $g_{el-ph}$, and $k_{ph-ph}$ that increases monotonically with the static temperature, as shown in SI Section S3. A temperature-independent $g_{el-ph}$ is line with the absence of electronic and structural phase transitions in the measured temperature range. On the other hand, $k_{ph-ph}$ is expected to vary linearly with temperature at high temperatures,[30,31] consistent with the results in Figure S3.

## 2.2 Disordering of itinerant spins by electron-phonon spin-flip scattering

Our UEDS measurements show that electron-phonon thermalization in MnBi$_2$Te$_4$ occurs primarily via optical phonons. Next, we measure the demagnetization associated with the electron-optical phonon scattering events. The phonon-mediated spin-flip scattering process that causes demagnetization is schematically shown in **Figure 3a**, where $a_{SF}$ is the spin-flip probability. We note that such a mechanism can act as a major channel of ultrafast demagnetization only in the presence of an itinerant spin subsystem, i. e. spin-polarized itinerant bands. The AFM order in MnBi$_2$Te$_4$ has previously been characterized as originating in localized Mn $3d^5$ spins. Our results, below, clearly show the existence of an itinerant spin subsystem consisting mainly of Bi and Te $p$ states, which disorder at ultrafast timescale via electron-phonon spin-flip processes.

In Figure 3b, we show the spin-polarized density of states projected onto atomic orbitals, calculated using density functional theory (DFT). Localized Mn $3d$ spins are clearly the dominant spin-polarized bands, however, a clear spin-splitting is also observed in the valence and conduction bands, which are due to Bi and Te states. While pristine MnBi$_2$Te$_4$ is an insulator, actual samples are metallic due to defect doping, with the Fermi level ($E_F$) at ~0.2 eV above the conduction band edge. The spin-splitting in the conduction band thus leads to a finite itinerant magnetic moment, which we calculate to be ~0.03 $\mu_B$ consisting almost entirely of Bi $5p$- and Te $4p$-like states (see SI Figure S4).

We measure ultrafast demagnetization associated with electron-phonon scattering by employing time-resolved magneto-optic Kerr effect (trMOKE) measurements (see Methods). Our measurements, with pump and probe energies of 1.55 eV and 1.2 eV respectively, are directly sensitive to optical transitions and spin-polarization in the valence and conduction band, as illustrated by the dashed box in Figure 3b.

Importantly, the pump does not excite occupied localized Mn *3d* states, which are at ~6 eV below $E_F$, so that only itinerant *p*-like spins are directly disordered by spin-flip processes. The measurements are carried out at 30 K and at a field of 1 T to induce a static magnetization (~0.3 $\mu_B$) and finite Kerr rotation θ. The transient pump-induced Kerr rotation Δθ (see SI Section S5 for Kerr ellipticity), plotted in the lower panel of Figure 3c using blue circles (normalized to the maximum demagnetization), shows an ultrafast demagnetization occurring at picosecond timescales. The demagnetization exhibits a characteristic shoulder at ~2 ps, which also corresponds to the timescale of electron-optical phonon thermalization. These observations are consistent with a disordering of itinerant spins due to electron-phonon spin-flip processes as we show below.

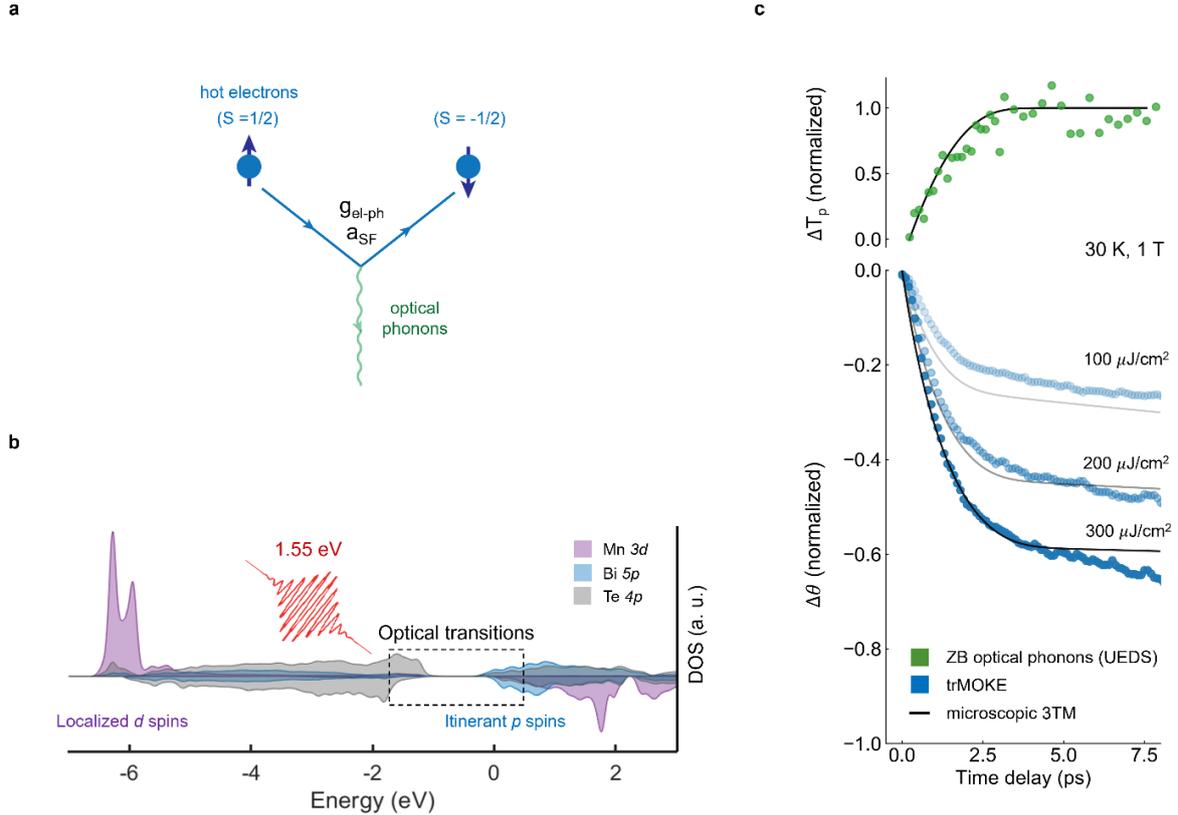

**Figure 3 Disordering of itinerant spins via electron-phonon spin-flip scattering. a,** Schematic of phonon-mediated spin-flip scattering, where *g* is the electron-phonon coupling constant, and $a_{SF}$ is the spin-flip probability. **b,** Density of states projected onto atomic orbitals. Bands with localized and itinerant spins are indicated. The 1.55 eV pump excitation is schematically shown. The region of optical transitions excited by the pump and detected by the probe is marked using a dashed box. **c,** (bottom) Results of trMOKE experiments (blue circles), with pump fluences labeled next to the plots. The black lines are fits to phenomenological model of electron-phonon spin-flip scattering shown schematically in panel (a). (top) Plot of normalized change in optical phonon temperature from UEDS measurements (green circles).

We model the experimental results in Figure 3c using a phenomenological model that considers individual electron-phonon spin-flip scattering events. To start with, we first consider only the itinerant *p* spin subsystem, ignoring its coupling to localized spins. Following the approach used in reference[32], we fix various model parameters using experimental inputs and DFT calculations (see SI Section S6). Additionally, we fix the electron-phonon scattering rate $g_{el-ph}$ to the value obtained from our UEDS measurements, leaving the spin-flip probability $a_{SF}$ and an effective exchange splitting as the only variables. The equations are solved numerically and fit to the entire fluence-dependent experimental dataset. The fit results are shown using solid black lines in Figure 3c. The following observations can be made – (i) the experimental results show good qualitative agreement with the model of electron-phonon spin-flip scattering. In particular, the temporal dynamics, including the fluence-independent shoulder at 2 ps, change in slope across the shoulder, and overall fluence-dependent scaling are captured by our model. (ii) A spin-flip probability $a_{SF}$ of 0.06 and an effective exchange splitting of 10 meV are obtained, both of which are physically reasonably values (see SI Section S6 for a discussion).[24] (iii) In addition to the timescale of demagnetization, the model predicts the dynamics of phonon populations due to electron-phonon thermalization, denoted in the top panel of Figure 3c using a solid black line. This prediction is in excellent agreement with the phonon population rate directly measured using UEDS, despite the substantially different temperatures and fluence used in the trMOKE experiments, validating our assumption of a temperature-independent electron-phonon scattering rate.

Our model of electron-phonon spin-flip scattering provides a good qualitative description of itinerant spin demagnetization at picosecond timescales. However, this simplified model is subject to some limitations. In particular, the model neglects the exchange coupling $J_{pd}$ between the itinerant Bi and Te *p* spins and localized Mn *3d* spins. Though trMOKE is a direct probe of magnetic contrast in optically accessible bands, it is additionally also indirectly sensitive to localized spins via the interatomic exchange coupling $J_{pd}$, so that the experimentally measured dynamics include contributions from both spin subsystems. In the following sections, we measure and quantify this exchange coupling, and capture its effect on the magnetic dynamics using atomistic simulations.

## 2.3 Demagnetization of localized Mn spins via $J_{pd}$

In order to isolate the dynamics of localized Mn spins, we employ time-resolved resonant soft X-ray scattering (RSXS) at the Mn $L_3$ edge (corresponding to *2p→3d* dipole transitions, see schematic in **Figure 4a**). The measurements were carried out at an angle of incidence of 22.5°, close to the Bragg reflection condition for the (0 0 1.5) peak that corresponds to the doubling of the unit cell with the onset of AFM order, to maximize the magnetic contrast (see Methods). The temperature-dependent static spectra plotted in Figure 4a show that reflectivity at the $L_3$ edge is sensitive to the AFM order (see SI Section S6 for static temperature-dependence). We carry out pump-probe measurements with a 1.2 eV pump (see Methods) that excites itinerant spins as in the trMOKE experiments, and the probe energy fixed at 638 eV to resonantly probe the antiferromagnetic order of localized Mn spins. In addition, these measurements are carried out at 0 T, and are thus a direct probe of the AFM order, rather than a polarized paramagnetic state as in the trMOKE measurements.

The time-resolved RSXS results are plotted in Figure 4b. Two distinct timescales are observed in the dynamics, with varying amplitudes as a function of fluence. An initial, fast demagnetization (i. e melting of antiferromagnetic order) occurs at timescales below the experimental resolution (probe pulsewidth ~90 ps), indicated using dashed lines. This is followed by a slow demagnetization occurring over hundreds of picoseconds. At sufficiently high fluences (~100 μJ/cm$^2$), the initial fast demagnetization fully melts the magnetic order. The different timescales and fluence-dependence are suggestive of two distinct mechanisms of demagnetization of localized Mn *3d* spins.

We first consider the slow process. This can be explained as the result of thermalization between the localized Mn *3d* spins and the lattice. Making use of a two-temperature model, separating the lattice and localized spin subsystems, the set of fluence-dependent data in the low fluence limit (i. e. where the initial fast demagnetization is negligible) can be fit (solid black lines in Figure 4b) to a single spin-lattice thermalization constant $g_{SL}$ (see SI Section S8). The thermalization process is illustrated by plotting the transient spin and lattice temperatures in the top panel of Figure 4b. While the lattice thermalizes within the first 15 ps (as shown in Figure 2d), the spins are only weakly coupled to the lattice, resulting in a thermalization timescale of the order of 400 ps. Unlike ultrafast demagnetization of itinerant spins via spin-flip scattering, spin-lattice thermalization does not require hot electrons. The disordering of spins instead occurs due to a coupling between the angular momentum of localized *3d* spins and thermal lattice motions. The coupling is mediated by the orbital angular momentum $L$ via the spin-orbit interaction $\xi L.S$. The orbital angular momentum is nominally fully quenched in the Mn *3d$^5$* spin state (i. e. $L = 0$), potentially resulting in a weak coupling of the lattice to Mn spin order, and thus the extremely long thermalization timescales. The weak spin-orbit interaction is also reflected in the small single-ion anisotropy $K_d$ of Mn *3d* spins.[12]

The pulsewidth-limited initial demagnetization is at odds with the above discussion. On the other hand, this timescale is consistent with demagnetization via spin-flip scattering observed in the conduction band. This suggests that the localized Mn *3d* spins are strongly coupled to itinerant Bi and Te *p* spins via an interatomic exchange interaction $J_{pd}$. In such a scenario, a sufficiently high-fluence pump excitation would completely disorder not just the itinerant *p* spins, but also the localized *3d* spins at timescales comparable to electron-optical phonon scattering. At lower fluences, only a partial disordering of localized *3d* spins would occur, allowing for subsequent demagnetization via the slower thermalization process. The results in Figure 4b are completely consistent with the above description. Such a picture of two distinct demagnetization processes with disparate timescales is further supported by temperature-dependent trMOKE measurements, discussed in SI Section S9. The fast initial demagnetization observed in the RSXS measurements is thus evidence for a strong interatomic exchange interaction $J_{pd}$ that disorders the Mn *3d* localized spins concurrently with the itinerant spins.

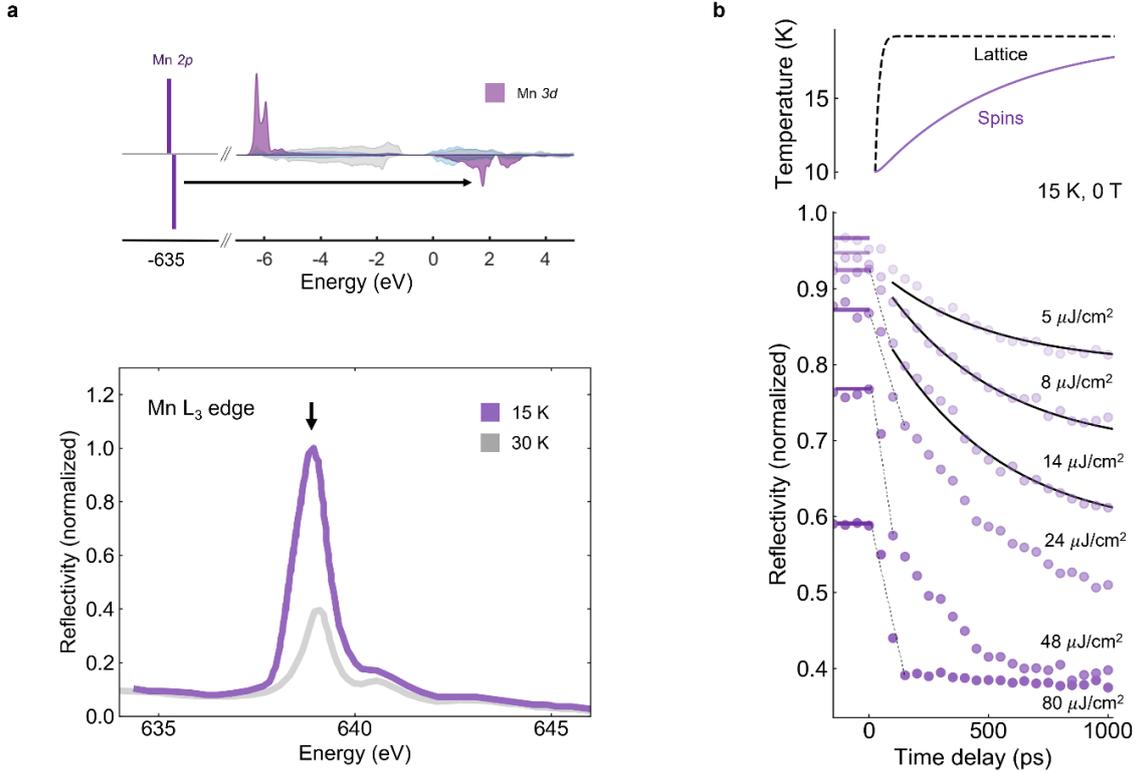

**Figure 4 Disordering of Mn *3d* localized spins via $J_{pd}$. a,** Schematic of resonant optical transition (top), and static resonant soft X-ray scattering (RSXS) spectrum (bottom). **b,** (bottom) RSXS results (purple circles) and fit (solid black line) to spin-lattice two-temperature model, with pump fluences labeled next to plots. (top) Spin and lattice temperatures at pump fluence of 20 $\mu$J/cm$^2$ extracted from two-temperature model.

**2.4 Estimating $J_{pd}$ using Landau-Lifshitz-Gilbert simulations**

In order to provide phenomenological insight into the experimentally observed coupled magnetic dynamics, we write down a Heisenberg-like Hamiltonian describing coupled *p* and *d* spins in MnBi$_2$Te$_4$ –

$$H = -\sum_{ij} J_d^{ij} S_d^i \cdot S_d^j - \sum_i K_d {S_d^{zi}}^2 - J_{pd} \sum_i S_p^i \cdot S_d^i \tag{1}$$

Here, $S_d^i$ and $S_p^i$ are Mn *3d* and Bi/Te *p*-like spins at site *i*, respectively. $J_d^{ij}$ is the exchange coupling constant of Mn *3d* spins including $J_{d,IP}$ for nearest-neighbor in-plane spins and $J_{d,OOP}$ for nearest-neighbor out-of-plane neighbors, $K_d$ is the single-ion anisotropy coefficient, $S_d^{zi}$ is the z-component of $S_d^i$, and $J_{pd}$ is the interatomic exchange interaction between Mn *3d* spins and Bi/Te *p* spins. Since $|S_p/S_d|$ ~ 0.01, we neglect terms of the order $S_p^2$.

Based on the above ratio of the magnitudes of *p* and *d* spins, $J_{pd}$ must have a value ~100 times larger than $J_d$ in order for the last term to have a significant contribution to the magnetic dynamics, as indicated by the concurrent *p-d* demagnetization observed in our measurements. Using $J_d$ = 0.12 meV extracted from inelastic neutron scattering,[12] we can estimate that $J_{pd}$ must be atleast ~10 meV.

To substantiate these arguments, we employ orbital-resolved atomistic Landau-Lifshitz-Gilbert simulations (see Methods for simulation details). Here nonequilibrium spin dynamics are determined by means of the spin Hamiltonian, along with orbital-resolved phenomenological 'damping' terms $\alpha_p$ and $\alpha_d$ that couple the *p* and *d* spin dynamics to the electronic and lattice heat baths, respectively. We use previously reported experimental results to fix the values of $J_{d,IP}$, $J_{d,OOP}$, $K_d$, and $S_d$, (see Section S10 for values) and estimate $S_p$ using density functional theory calculations (see Section S4).[10,12] We find good agreement between the simulated Mn *3d* spin dynamics and the experimental RSXS results using $\alpha_d$ = 0.003 and $\alpha_p$ = 0.5, with $J_{pd}$ in the range of 10-50 meV.

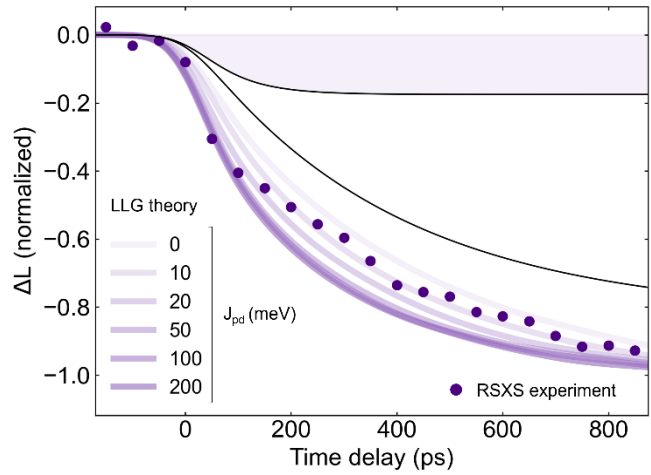

**Figure 5 Estimating $J_{pd}$ using LLG simulations.** The pump-induced change in the antiferromagnetic order parameter *L* corresponding to Mn *3d* spins is plotted, with circles denoting results from resonant soft X-ray scattering (RSXS) measurements at a pump fluence of 29 µJ/cm², and solid purple lines denoting results of Landau-Lifshitz-Gilbert (LLG) simulations as a function of the interatomic exchange coupling $J_{pd}$. The black lines are the individual exponential decay functions used to fit the simulation results for $J_{pd}$ = 20 meV. The purple region highlights the disordering of spins achieved due to $J_{pd}$. The simulation results are convoluted with a 90 ps Gaussian to model the experimental probe pulsewidth.

In particular, the simulations accurately reproduce the static ordering temperature $T_N$ = 24 K (see Figure S8a), and in the limit of $J_{pd}$ = 0, reproduce the low fluence spin-lattice thermalization dynamics (see Figure S8b). In **Figure 5**, we compare the experimental Mn *3d* demagnetization at an intermediate fluence (29 µJ/cm²) with the results of our LLG simulations for various values of $J_{pd}$. The results show that a value of $J_{pd}$ >10 meV is necessary to reproduce the experimentally observed initial fast demagnetization step. The simulated dynamics are well-described by a bi-exponential decay function, a representative sample of which is shown for $J_{pd}$ = 20 meV. The dynamics consist of a slow step (black line) with a time constant of ~380 ps corresponding to spin-lattice thermalization (also see Figure S8b), and a fast step with a time constant of ~70 ps (black line with purple shading) corresponding to the disordering of Mn *3d* spins due to its coupling to itinerant spins via $J_{pd}$. The time constant of the fast step is consistent with the pulsewidth-limited disordering observed in our experiments.

## 3. Conclusion

The transfer of spin angular momentum is a critical aspect of ultrafast magnetic dynamics. Our work demonstrates the role played by optical phonons in this process through direct measurement of both nonequilibrium phonon and spin dynamics. The momentum- and mode-resolved information in our UEDS experiments goes beyond simple phenomenological approaches to electron-phonon coupling and reveals the rich temporal dynamics of nonequilibrium phonons in complex materials such as $MnBi_2Te_4$. In particular, UEDS is a direct probe of electron-optical phonon scattering, improving upon mode-integrated probes such as ultrafast diffraction which only provide a lattice thermalization time constant averaged over all phonon modes.

Our work illustrates that with a sufficiently large $J_{pd}$, not only do the miniscule itinerant $p$-like magnetic moments (~0.03 $\mu_B$) follow the dynamics of dominant localized Mn $3d$ magnetic moments (~5 $\mu_B$), but that the reverse is also possible, i. e. disordering of the $p$-like spins results in a concurrent disordering of the localized $3d$ spins down to picosecond timescales. We note that this is not always the case. For example, in Gd, an elemental ferromagnet, localized $4f$ spins (7 $\mu_B$) and itinerant $5d$ spins (0.55 $\mu_B$) exhibit disparate ultrafast demagnetization timescales, despite a comparably large exchange coupling of 130 meV.[33] On the other hand, in dilute magnetic semiconductors (DMS), localized $d$ spins that are isolated from the initial carrier excitation are found to disorder at timescales comparable to carrier-lattice thermalization,[34–36] notwithstanding the vanishing itinerant magnetic moment, similar to the spin dynamics in $MnBi_2Te_4$ reported here. This disparity may possibly arise from qualitative differences in chemistry – rare-earth elemental ferromagnets such as Gd have localized spins in $f$ orbitals, whereas DMS systems and $MnBi_2Te_4$ have like $d$-like localized spins. The striking resemblance between spin dynamics in DMS systems and $MnBi_2Te_4$ points to a similar mechanism stabilizing long-range magnetic order in these materials. In particular, in addition to superexchange, an RKKY-like exchange coupling may exist between localized spins in $MnBi_2Te_4$, mediated by carriers, much like in DMS systems.[16,17] Anomalies observed in magneto-transport measurements and the dramatic variation in magnetic properties with Fermi level tuning by chemical doping[11,37] are indeed consistent with this hypothesis.

Quantized transport phenomena in magnetic topological materials are the consequence of gapping of the topological surface states by localized magnetic order, which is enabled by the exchange coupling $J_{pd}$. This interaction has to date only been probed and quantified in topological insulators randomly doped with magnetic ions.[16,38] Our work quantifies this in an intrinsic magnetic topological material, by direct measurement of itinerant $p$ and localized $3d$ spin dynamics in $MnBi_2Te_4$. Assuming a fully saturated localized magnetic moment of $S = 5/2$, A-type antiferromagnetic order at the surface, and $J_{pd}$ >10 meV, we predict an exchange gap in the topological surface states of >25 meV, which is in good agreement with theoretical estimates in the range of 30-60 meV.[19,39] We also note that 25 meV is well above the energy-resolution of cutting-edge ARPES tools. The lack of an observable exchange gap in ARPES measurements on $MnBi_2Te_4$ thus cannot be explained by a weak $J_{pd}$ coupling as proposed in the literature.[14] Surface-sensitive magnetic measurements indicate that the A-type AFM order persists into the surface layers,[14,21] and magnetic domains in $MnBi_2Te_4$ were measured to have dimensions of tens of microns,[15] likely ruling out surface spin reconstruction and nanometer-scale domains as possible reasons for the vanishing exchange gap. An alternative explanation is that the topological surface states extend into the bulk, either due to a large decay length,[20,21] or surface-bulk hybridization.[40] In such a scenario, since the interlayer magnetic order is AFM, the effective localized magnetic moment would be substantially lowered, resulting in a vanishingly small exchange gap. Our results support such an explanation. It has been recently reported that $MnBi_8Te_{13}$, which has a ferromagnetic ground state and thus a robust net magnetization, exhibits a saturated exchange gap of 33 meV as measured by ARPES,[41] in excellent agreement with our estimation. The strong exchange coupling between localized spins and itinerant bands observed in our experiments thus validates the strategy of using localized magnetic order to create magnetic topological phases in the $Mn(Bi,Sb)_{2n}Te_{3n+1}$ family of materials. More generally, our study demonstrates how multimodal ultrafast approaches may be used to quantify exchange couplings in complex functional quantum materials interfaced with magnetism.

**Methods**

**Crystal growth and characterization**

Single crystals of MnBi$_2$Te$_4$ were grown using a self-flux method as reported elsewhere. X-ray diffraction was used to check the crystallinity and phase of the crystals. SQUID magnetometry was used to confirm the antiferromagnetic order with a Néel temperature of 24 K.

**Ultrafast electron diffuse scattering**

Ultrafast electron scattering measurements were carried out at the MeV-UED beamline at the SLAC National Laboratory. The technical details of the experimental setup and principles are explained elsewhere.[42,43] The sample was excited using 800 nm (1.55 eV), 60 fs pulses, at a fluence of 5 mJ/cm$^2$. The high fluence is necessary to produce a sufficiently large pump-induced change in electron scattering. No laser-induced damage was observed in fluence-dependent damage studies, and the signal was repeatable over thousands of cycles. The kinetic energy of the electron bunches is 3.7 MeV and the pulsewidth is ~100 fs. The pump and probe spot sizes were 464×694 μm and ~70 μm.

Measurements were carried out on flakes of average thickness of around 100 nm, exfoliated using an ex-situ transfer stage onto an amorphous Si$_3$N$_4$ membrane. The flakes were protected with an additional layer of amorphous Si$_3$N$_4$ to prevent degradation.

Ultrafast electron diffuse scattering (UEDS) intensities were obtained by averaging over several scans, with individual electron scattering images normalized and aligned to account for electron beam charge and pointing fluctuations, respectively. Time-resolved diffuse scattering intensities were obtained by integrating over appropriate regions of integration and averaged over symmetry-related points in the scattering phase space based on the *R*-3*m* space group of MnBi$_2$Te$_4$. The pump-induced changes in diffuse scattering intensity were obtained by subtracting diffuse scattering intensity integrated over time delays $t < 0$ from the time-resolved intensity. Regions with negative pump-induced changes in scattering intensity corresponding to the tails of Bragg peaks were masked using a Heaviside function, to avoid interference with the diffuse scattering.

**Time-resolved magneto-optic Kerr rotation**

Time-resolved magneto-optical Kerr effect (MOKE) measurements were carried out using a 1040 nm 200 kHz Spectra-Physics Spirit Yb-based hybrid-fiber laser coupled to a noncollinear optical parametric amplifier. The amplifier produces <50 fs pulses centered at 800 nm (1.55 eV), used as the pump beam. The 1040 nm (1.2 eV) output is converted to white light, centered at 1025 nm with a FWHM of 20 nm, by focusing it inside a YAG (Yttrium Aluminum Garnet) crystal. The white light is subsequently compressed to ~50 fs pulses using a prism compressor pair and used as the probe beam. The pump and the probe beams are aligned to propagate along the [001] axis of the crystal, at near normal incidence. A Wollaston prism is used to separate S- and P-polarized probe beam reflected off the sample surface. The photoinduced Kerr rotation was measured using balanced photodiodes. Details of the Kerr rotation analysis are outlined elsewhere.[44]

The samples were placed in a magneto-optical closed-cycle cryostat (Quantum Design OptiCool). Measurements were carried out at a magnetic field of 1 T applied normal to the sample surface (along the [001] direction). The sample temperature was varied from 10 K to 50 K. A pump fluence of ~100 μJ/cm$^2$ was used.

**Time-resolved resonant soft X-ray scattering**

Time-resolved resonant soft X-ray scattering measurements were carried out at beamline 4-ID-C at the Advanced Photon Source in Argonne National Laboratory. Optical pulses centered at 1030 nm (1.2 eV) with a pulsewidth of 300 fs at a repetition rate of 108 kHz were used as the pump. Circularly polarized soft X-ray pulses of variable energy (400 – 2500 eV) with a pulsewidth of ~90 ps at a repetition rate of 6.5 MHz were used as the probe. The pump beam was incident at near normal incidence. The probe beam was incident at an angle of 22.5º in a reflection geometry, and the reflected beam was detected using an avalanche photodiode.

The samples were mounted on a flow cryostat in vacuum and measurements were carried out at a temperature of 15 K at zero magnetic field.

**Electronic structure calculations**

Density functional theory calculations were carried out using the Vienna Ab Initio Simulation Package (VASP)[45–49] with the PBE exchange correlation functional[50] and van der Waals correction via the DFT-D3[51,52] method with Becke-Jonson damping. A Hubbard $U$ was also added to the Mn (4 eV) using Dudarev's[53] approach. A non-primitive cell containing two Mn atoms was used to obtain the equilibrium geometry of the system with AFM-A magnetic structure. An energy cutoff of 300 eV was used for all calculations. A 4×4×4 Γ-centered $k$-point mesh was used for equilibrium relaxations. The general energy convergence threshold was $1\times10^{-8}$ eV and the force convergence threshold for relaxation was $1\times10^{-5}$ eV/Å. Gaussian smearing with a 0.02 eV width was also used in all relaxation and single-point energy calculations. Spin-polarized density of states calculations were done assuming a FM magnetic structure and employed the tetrahedron method. Supercells for magnetic exchange calculations were generated using VESTA.[54]

**Landau-Lifshitz-Gilbert simulations**

We consider a system consisting of 1,024,000 sites each possessing a primitive unit cell with a formula of MnBi$_2$Te$_4$. We do not distinguish the atoms within a unit cell for simplicity. We simulate the Langevin dynamics of the spins, where the evolution of the localized $d$ spin $S_d^i$ and itinerant $p$ spin $S_p^i$ of the site $i$ follow the stochastic LLG equations, i. e.,

$$\frac{dS_d^i}{dt} = -\frac{\gamma_0}{(1+\alpha_d^2)S_d}\left(S_d^i \times H_d^i + \frac{\alpha_d}{S_d}S_d^i \times (S_d^i \times H_d^i)\right) \tag{2}$$

$$\frac{dS_p^i}{dt} = -\frac{\gamma_0}{(1+\alpha_p^2)S_p}\left(S_p^i \times H_p^i + \frac{\alpha_p}{S_p}S_p^i \times (S_p^i \times H_p^i)\right) \tag{3}$$

Here $t$ is the time, $\gamma_0$ is the electron gyromagnetic ratio, $S_d^i$ and $S_p^i$ are the saturated magnitudes of $d$ spin and $p$ spin of site $i$, respectively, and $\alpha_d$ and $\alpha_p$ are their damping constants. The effective fields $H_d^i$ and $H_p^i$ include driving forces from Hamiltonian $H$ and thermal fluctuations given by white-noise terms, i. e. $H_d^i = -\partial H/\partial S_d^i + H_{d,noise}^i$ and $H_d^i = -\partial H/\partial S_p^i + H_{p,noise}^i$.

The Hamiltonian $H$ is as shown in Equation 1. For simplicity, we only consider a nonzero $J_d^{ij}$ between a site and each of its 6 nearest in-plane neighbors ($J_{d,IP}$) as well as 6 nearest out-of-plane neighbors ($J_{d,OOP}$). $J_{d,IP} > 0$ together with $J_{d,OOP} < 0$ will give rise to an A-type antiferromagnetic order of the spins at low temperatures. Exchange interactions between $p$ spins of different sites and that between $d$ and $p$ spins of different sites are neglected.

The white-noise terms $H_{d,noise}^i$ and $H_{p,noise}^i$ are random vectors that fulfill[55]

$$\langle H_{d,noise}^i(t) \rangle = 0, \quad \langle H_{d,noise,m}^i(t_1), H_{d,noise,n}^j(t_2) \rangle - \delta_{ij}\delta_{mn}\delta(t_1-t_2)2\alpha_d k_B T_l S_d/\gamma_0$$

$$\langle H_{p,noise}^i(t) \rangle = 0, \quad \langle H_{p,noise,m}^i(t_1), H_{p,noise,n}^j(t_2) \rangle - \delta_{ij}\delta_{mn}\delta(t_1-t_2)2\alpha_p k_B T_e S_p/\gamma_0$$

$$\langle H_{d,noise,m}^i(t_1), H_{p,noise,n}^j(t_2) \rangle = 0 \tag{4}$$

Here subscripts $m, n = 1, 2, 3$ indicate the coordinates of a 3-dimensional Cartesian coordinate system, $k_B$ is the Boltzmann constant, $T_l$ and $T_e$ are the lattice and electronic temperatures of a two-temperature model for the lattice and electrons, as determined using the el-ph thermalization time from UEDS measurements.

The material constants used in the simulation are $S_d = 2.5$, $S_p = 0.025$, $\alpha_d = 0.003$, $\alpha_p = 0.5$, $\gamma_0 = 1.761 \times 10^{11}$ T$^{-1}$s$^{-1}$, $J_{d,IP} = 0.16$ meV, $J_{d,OOP} = -0.022$ meV, $K_d = 0.05$ meV, and $J_{pd} = 10$-$200$ meV (see SI Section S10 for details)

**Acknowledgements**

H.P., V.A.S., H.W., P.K., M.P., N.Z.K., A.M.L., R.D.A., J.M.R., J.W.F., and V.G. acknowledge support from the DOE-BES grant DE-SC0012375. H.P., T.Y., L-Q.C., and V.G. acknowledge support from the DOE Computational Materials program, DE-SC0020145. Support for crystal growth and characterization was provided by the National Science Foundation through the Penn State 2D Crystal Consortium-Materials Innovation Platform (2DCC-MIP) under NSF cooperative agreement DMR-1539916. D.P. was supported by the Army Research Office (ARO) under grant no. W911NF-15-1-0017. SLAC MeV-UED is supported in part by the DOE BES SUF Division Accelerator & Detector R&D program, the LCLS Facility, and SLAC under Contract Nos. DE-AC02–05-CH11231 and DE-AC02–76SF00515.

H.P., V.A.S, and V.G. conceived the project. H.P., V.A.S., H.W., X.S., A.H.R., M-F.L, S.P., A.M.L., V.G, and X.W. carried out the ultrafast electron diffuse scattering measurements at the SLAC National Laboratory. P.K., M.P., H.P., R.A., and V.G. carried out the magneto-optic Kerr rotation measurements. V.A.S and J.W.F carried out the resonant soft X-ray scattering measurements in Argonne National Laboratory. S-H.L and Z.M. synthesized single crystals of MnBi$_2$Te$_4$ used in the measurements. L.M. and H.P. carried out the SQUID magnetometry measurements. T.Y. and L-Q.C. carried out the LLG simulations. N.K, D.P., and J.M.R carried out the density functional theory simulations. H.P. analyzed the experimental results with inputs from all authors. H.P. wrote the manuscript with inputs from all authors.

The authors declare no competing interests.

[1] M. M. Otrokov, I. I. Klimovskikh, H. Bentmann, D. Estyunin, A. Zeugner, Z. S. Aliev, S. Gaß, A. U. B. Wolter, A. V. Koroleva, A. M. Shikin, M. Blanco-Rey, M. Hoffmann, I. P. Rusinov, A. Y. Vyazovskaya, S. V. Eremeev, Y. M. Koroteev, V. M. Kuznetsov, F. Freyse, J. Sánchez-Barriga, I. R. Amiraslanov, M. B. Babanly, N. T. Mamedov, N. A. Abdullayev, V. N. Zverev, A. Alfonsov, V. Kataev, B. Büchner, E. F. Schwier, S. Kumar, A. Kimura, L. Petaccia, G. Di Santo, R. C. Vidal, S. Schatz, K. Kißner, M. Ünzelmann, C. H. Min, S. Moser, T. R. F. Peixoto, F. Reinert, A. Ernst, P. M. Echenique, A. Isaeva, E. V. Chulkov, *Nature* **2019**, *576*, 416.
[2] Y. Deng, Y. Yu, M. Z. Shi, Z. Guo, Z. Xu, J. Wang, X. H. Chen, Y. Zhang, *Science (80-. ).* **2020**, *367*, 895.
[3] C. Liu, Y. Wang, H. Li, Y. Wu, Y. Li, J. Li, K. He, Y. Xu, J. Zhang, Y. Wang, *Nat. Mater.* **2020**, *19*, 522.
[4] S. H. Lee, D. Graf, L. Min, Y. Zhu, H. Yi, S. Ciocys, Y. Wang, E. S. Choi, R. Basnet, A. Fereidouni, A. Wegner, Y. F. Zhao, K. Verlinde, J. He, R. Redwing, V. Gopalan, H. O. H. Churchill, A. Lanzara, N. Samarth, C. Z. Chang, J. Hu, Z. Q. Mao, *Phys. Rev. X* **2021**, *11*, 31032.
[5] Y. Tokura, K. Yasuda, A. Tsukazaki, *Nat. Rev. Phys.* **2019**, *1*, 126.
[6] H. Zhang, C.-X. Liu, X.-L. Qi, X. Dai, Z. Fang, S.-C. Zhang, *Nat. Phys.* **2009**, *5*, 438.
[7] A. S. Núñez, J. Fernández-Rossier, *Solid State Commun.* **2012**, *152*, 403.
[8] Q. Liu, C.-X. Liu, C. Xu, X.-L. Qi, S.-C. Zhang, *Phys. Rev. Lett.* **2009**, *102*, 156603.
[9] R. Yu, W. Zhang, H.-J. Zhang, S.-C. Zhang, X. Dai, Z. Fang, *Science (80-. ).* **2010**, *329*, 61.
[10] J.-Q. Yan, Q. Zhang, T. Heitmann, Z. Huang, K. Y. Chen, J.-G. Cheng, W. Wu, D. Vaknin, B. C. Sales, R. J. McQueeney, *Phys. Rev. Mater.* **2019**, *3*, 064202.
[11] S. H. Lee, Y. Zhu, Y. Wang, L. Miao, T. Pillsbury, H. Yi, S. Kempinger, J. Hu, C. A. Heikes, P. Quarterman, W. Ratcliff, J. A. Borchers, H. Zhang, X. Ke, D. Graf, N. Alem, C.-Z. Chang,

N. Samarth, Z. Mao, *Phys. Rev. Res.* **2019**, *1*, 12011.

[12] B. Li, J.-Q. Yan, D. M. Pajerowski, E. Gordon, A.-M. Nedić, Y. Sizyuk, L. Ke, P. P. Orth, D. Vaknin, R. J. McQueeney, *Phys. Rev. Lett.* **2020**, *124*, 167204.

[13] B. Li, D. M. Pajerowski, S. Riberolles, L. Ke, J. Q. Yan, R. J. McQueeney, *Phys. Rev. B* **2020**, *104*, L220402.

[14] P. M. Sass, J. Kim, D. Vanderbilt, J. Yan, W. Wu, *Phys. Rev. Lett.* **2020**, *125*, 037201.

[15] P. M. Sass, W. Ge, J. Yan, D. Obeysekera, J. J. Yang, W. Wu, *Nano Lett.* **2020**, *20*, 2609.

[16] J. S. Dyck, P. Hájek, P. Lošt'ák, C. Uher, *Phys. Rev. B* **2002**, *65*, 115212.

[17] F. Matsukura, H. Ohno, A. Shen, Y. Sugawara, *Phys. Rev. B* **1998**, *57*, R2037.

[18] M. Köpf, J. Ebad-Allah, S. H. Lee, Z. Q. Mao, C. A. Kuntscher, *Phys. Rev. B* **2020**, *102*, 165139.

[19] Y.-J. Hao, P. Liu, Y. Feng, X.-M. Ma, E. F. Schwier, M. Arita, S. Kumar, C. Hu, R. Lu, M. Zeng, Y. Wang, Z. Hao, H.-Y. Sun, K. Zhang, J. Mei, N. Ni, L. Wu, K. Shimada, C. Chen, Q. Liu, C. Liu, *Phys. Rev. X* **2019**, *9*, 041038.

[20] Y. J. Chen, L. X. Xu, J. H. Li, Y. W. Li, H. Y. Wang, C. F. Zhang, H. Li, Y. Wu, A. J. Liang, C. Chen, S. W. Jung, C. Cacho, Y. H. Mao, S. Liu, M. X. Wang, Y. F. Guo, Y. Xu, Z. K. Liu, L. X. Yang, Y. L. Chen, *Phys. Rev. X* **2019**, *9*, 041040.

[21] D. Nevola, H. X. Li, J.-Q. Yan, R. G. Moore, H.-N. Lee, H. Miao, P. D. Johnson, *Phys. Rev. Lett.* **2020**, *125*, 117205.

[22] R. J. Elliott, *Phys. Rev.* **1954**, *96*, 266.

[23] Y. Yafet, in *Solid State Phys.*, **1963**, pp. 1–98.

[24] B. Koopmans, G. Malinowski, F. Dalla Longa, D. Steiauf, M. Fähnle, T. Roth, M. Cinchetti, M. Aeschlimann, *Nat. Mater.* **2010**, *9*, 259.

[25] R. Xu, T. C. Chiang, *Zeitschrift fur Krist.* **2005**, *220*, 1009.

[26] J. Wu, F. Liu, M. Sasase, K. Ienaga, Y. Obata, R. Yukawa, K. Horiba, H. Kumigashira, S. Okuma, T. Inoshita, H. Hosono, *Sci. Adv.* **2019**, *5*, eaax9989.

[27] J. A. Tomko, S. Kumar, R. Sundararaman, P. E. Hopkins, *J. Appl. Phys.* **2021**, *129*, 193104.

[28] J. K. Chen, W. P. Latham, J. E. Beraun, *J. Laser Appl.* **2005**, *17*, 63.

[29] Z. Lin, L. V. Zhigilei, V. Celli, *Phys. Rev. B* **2008**, *77*, 075133.

[30] P. G. Klemens, *Phys. Rev.* **1966**, *148*, 845.

[31] T. Konstantinova, J. D. Rameau, A. H. Reid, O. Abdurazakov, L. Wu, R. Li, X. Shen, G. Gu, Y. Huang, L. Rettig, I. Avigo, M. Ligges, J. K. Freericks, A. F. Kemper, H. A. Dürr, U. Bovensiepen, P. D. Johnson, X. Wang, Y. Zhu, *Sci. Adv.* **2018**, *4*, eaap7427.

[32] D. Longa, *Laser-Induced Magnetization Dynamics : An Ultrafast Journey among Spins and Light Pulses*, **2008**.

[33] B. Frietsch, J. Bowlan, R. Carley, M. Teichmann, S. Wienholdt, D. Hinzke, U. Nowak, K. Carva, P. M. Oppeneer, M. Weinelt, *Nat. Commun.* **2015**, *6*, 8262.

[34] J. Wang, C. Sun, Y. Hashimoto, J. Kono, G. A. Khodaparast, Ł. Cywiński, L. J. Sham, G. D. Sanders, C. J. Stanton, H. Munekata, *J. Phys. Condens. Matter* **2006**, *18*, R501.

[35] J. Wang, C. Sun, J. Kono, A. Oiwa, H. Munekata, Ł. Cywiński, L. J. Sham, *Phys. Rev. Lett.* **2005**, *95*, 167401.

[36] J. Wang, Ł. Cywiński, C. Sun, J. Kono, H. Munekata, L. J. Sham, *Phys. Rev. B* **2008**, *77*, 235308.

[37] J.-Q. Yan, S. Okamoto, M. A. McGuire, A. F. May, R. J. McQueeney, B. C. Sales, *Phys. Rev. B* **2019**, *100*, 104409.

[38] M. Ye, W. Li, S. Zhu, Y. Takeda, Y. Saitoh, J. Wang, H. Pan, M. Nurmamat, K. Sumida, F. Ji, Z. Liu, H. Yang, Z. Liu, D. Shen, A. Kimura, S. Qiao, X. Xie, *Nat. Commun.* **2015**, *6*, 8913.

[39] J. Li, Y. Li, S. Du, Z. Wang, B.-L. Gu, S.-C. Zhang, K. He, W. Duan, Y. Xu, *Sci. Adv.* **2019**, *5*, eaaw5685.

[40] X.-M. Ma, Z. Chen, E. F. Schwier, Y. Zhang, Y.-J. Hao, S. Kumar, R. Lu, J. Shao, Y. Jin, M. Zeng, X.-R. Liu, Z. Hao, K. Zhang, W. Mansuer, C. Song, Y. Wang, B. Zhao, C. Liu, K. Deng, J. Mei, K. Shimada, Y. Zhao, X. Zhou, B. Shen, W. Huang, C. Liu, H. Xu, C. Chen, *Phys. Rev. B* **2020**, *102*, 245136.

[41] R. Lu, H. Sun, S. Kumar, Y. Wang, M. Gu, M. Zeng, Y.-J. Hao, J. Li, J. Shao, X.-M. Ma, Z. Hao, K. Zhang, W. Mansuer, J. Mei, Y. Zhao, C. Liu, K. Deng, W. Huang, B. Shen, K. Shimada, E. F. Schwier, C. Liu, Q. Liu, C. Chen, *Phys. Rev. X* **2021**, *11*, 011039.

[42]  S. P. Weathersby, G. Brown, M. Centurion, T. F. Chase, R. Coffee, J. Corbett, J. P. Eichner, J. C. Frisch, A. R. Fry, M. Gühr, N. Hartmann, C. Hast, R. Hettel, R. K. Jobe, E. N. Jongewaard, J. R. Lewandowski, R. K. Li, A. M. Lindenberg, I. Makasyuk, J. E. May, D. McCormick, M. N. Nguyen, A. H. Reid, X. Shen, K. Sokolowski-Tinten, T. Vecchione, S. L. Vetter, J. Wu, J. Yang, H. A. Dürr, X. J. Wang, *Rev. Sci. Instrum.* **2015**, *86*, 073702.
[43]  X. Shen, R. K. Li, U. Lundström, T. J. Lane, A. H. Reid, S. P. Weathersby, X. J. Wang, *Ultramicroscopy* **2018**, *184*, 172.
[44]  D. J. Lovinger, E. Zoghlin, P. Kissin, G. Ahn, K. Ahadi, P. Kim, M. Poore, S. Stemmer, S. J. Moon, S. D. Wilson, R. D. Averitt, *Phys. Rev. B* **2020**, *102*, 085138.
[45]  G. Kresse, D. Joubert, *Phys. Rev. B* **1999**, *59*, 1758.
[46]  G. Kresse, J. Furthmüller, *Phys. Rev. B* **1996**, *54*, 11169.
[47]  G. Kresse, J. Hafner, *Phys. Rev. B* **1993**, *47*, 558.
[48]  G. Kresse, J. Furthmüller, *Comput. Mater. Sci.* **1996**, *6*, 15.
[49]  G. Kresse, J. Hafner, *Phys. Rev. B* **1994**, *49*, 14251.
[50]  J. P. Perdew, K. Burke, M. Ernzerhof, *Phys. Rev. Lett.* **1996**, *77*, 3865.
[51]  S. Grimme, J. Antony, S. Ehrlich, H. Krieg, *J. Chem. Phys.* **2010**, *132*, 154104.
[52]  S. Grimme, *J. Comput. Chem.* **2006**, *27*, 1787.
[53]  S. Dudarev, G. Botton, *Phys. Rev. B* **1998**, *57*, 1505.
[54]  K. Momma, F. Izumi, *J. Appl. Crystallogr.* **2011**, *44*, 1272.
[55]  U. Nowak, *Handbook of Magnetism and Advanced Magnetic Materials, Vol. 2*, Wiley, **2007**.

Supporting Information

**Large Itinerant Electron Exchange Coupling in the Magnetic Topological Insulator MnBi$_2$Te$_4$**

## S1. Ultrafast electron scattering experiments and analysis

Ultrafast electron scattering measurements were carried out as outlined in the Methods section. In general, electron scattering off a single crystal produces images with well-defined peaks corresponding to Bragg reflections (referred to as electron diffraction), and diffuse intensity corresponding to inelastic scattering events (referred to as electron diffuse scattering), in this case primarily scattering with phonons. The time-resolved experiments thus provide access to ultrafast diffraction as well as diffuse scattering intensities.

A representative static electron diffraction image is shown in Fig. S1a, with selected Bragg peaks labeled using white arrows. The ultrafast diffraction intensities provide information on nonequilibrium lattice thermalization timescales integrated over momentum and all phonon modes. These results have been discussed elsewhere[1]. Here, we instead focus on the ultrafast electron diffuse scattering (UEDS).

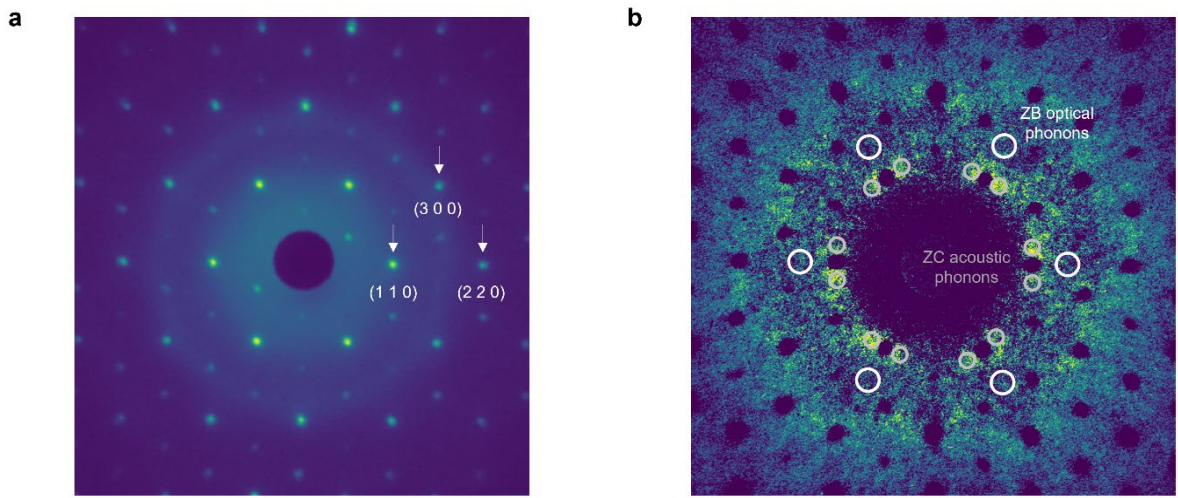

**Fig. S1 Ultrafast electron diffraction and diffuse scattering. a,** Representative static electron diffraction image. Selected Bragg peaks are indicated using white arrows and labeled. **b,** Representative ultrafast electron diffuse scattering, obtained by subtracting the intensity integrated over time delays $t < 0$ from the intensity integrated over time delays $t > 0$. Symmetry-related regions of integration corresponding to zone boundary (ZB) optical phonons (white circles) and zone-center acoustic phonons (grey circles) are shown and labeled.

A representative UEDS image is shown in Fig. S1b, obtained by subtracting the intensity integrated over time delays $t < 0$ (i. e. before pump excitation) from the intensity integrated over time delays $t > 0$. The symmetry-related regions of integration (ROIs) used to extract the data plotted in Fig. 1d are shown using circles. The ROIs within white circles correspond to zone-boundary (ZB) optical phonons, and those within grey circles correspond to zone-center (ZC) acoustic phonons.

## S2. Simulation of phonon structure factors

Diffuse scattering due to phonons may be described using the kinematical approximation[2]. The first-order diffuse scattering intensity is given by –

$$I_1(\boldsymbol{Q}) \propto \sum_i \frac{n_i(\boldsymbol{q})+1/2}{\omega_i(\boldsymbol{q})} |F_i(\boldsymbol{Q})|^2, \qquad (1)$$

where $\boldsymbol{Q}$ is the scattering wavevector, $\boldsymbol{q}$ is the reduced wavevector within the appropriate Brillouin zone, $i$ is an index running over all phonon branches, $n_i(\boldsymbol{q})$ is the population of phonon branch $i$ at momentum $\boldsymbol{q}$, $\omega_i(\boldsymbol{q})$ is the frequency of phonon branch $i$ at momentum $\boldsymbol{q}$, and $F_i(\boldsymbol{Q})$ is the one-phonon structure factor associated with phonon branch $i$.

The one-phonon structure factor is given by –

$$F_i(\boldsymbol{Q}) = \sum_s e^{-W_s Q^2} \frac{f_s(\boldsymbol{Q})}{\sqrt{\mu_s}} \left(\boldsymbol{Q} \cdot \boldsymbol{e}_{i,s}(\boldsymbol{q})\right), \qquad (2)$$

where $s$ is an index running over all the atoms in the unit cell, $W_s$ is an atom-dependent Debye-Waller factor, $f_s(\boldsymbol{Q})$ is the atomic structure factor for electron scattering, $\mu_s$ is the atomic mass, and $\boldsymbol{e}_{i,s}(\boldsymbol{q})$ is the eigendisplacement of atom $s$ in phonon branch $i$ at momentum $\boldsymbol{q}$.

For time-resolved diffuse scattering experiments as in our work, all of the above quantities are functions of time delay after pump excitation. In practice, it is assumed that the phonon frequencies and eigendisplacements remain unchanged, and that only the phonon populations $n_i(\boldsymbol{q})$ change with time. Our experiments thus provide access to ultrafast phonon populations. An explicit calculation of $F_i(\boldsymbol{Q})$ requires the calculation of eigendisplacements of each individual phonon branch resolved across the entire Brillouin zone. While this may be done in principle using density functional theory simulations, the large number of phonons (21) and their entwined dispersions make it challenging. Furthermore, even if $F_i(\boldsymbol{Q})$ was explicitly evaluated for all the phonon branches, resolving each of their individual contributions to the experimental diffuse scattering intensities would be practically impossible. Such approaches have previously been implemented successfully only in simple systems consisting of one or two atoms per unit cell[3–5].

We instead employ simplifying assumptions that allow us to resolve the diffuse scattering intensity into contributions from optical and acoustic phonon subsystems. First, we note from Equation 1 that the contribution due to acoustic phonons diverges at zone centers, where $\omega_i(\boldsymbol{q})$ vanishes. This suggests that the diffuse scattering intensity will be dominated by zone center acoustic phonons.

Next, given their divergence near zone centers, acoustic phonon structure factors may be approximated as coming only from near the zone center. This allows us to approximate the acoustic phonon branches by a toy model with linear dispersions, and eigendisplacements that are purely transverse or longitudinal, assumptions which are strictly valid only at the zone center. We can thus simulate the two in-plane acoustic phonon branches, namely transverse and longitudinal. The results, obtained by employing the above approximations in Equations 1 and 2, and assuming that the branches are populated thermally at a temperature of 300 K, are shown in Fig. S2.

On the other hand, for optical phonons, given their flat dispersion relations, Equation 1 implies that their contributions to the diffuse scattering intensity are relatively uniform across the scattering phase space, with no divergences, contrary to acoustic phonons. This insight, in combination with the experimental results, allows us to identify different regions in the scattering phase space that are dominated by acoustic and optical phonons respectively. The experimental diffuse scattering in Fig. 2b shows no signatures of the longitudinal acoustic phonon branch. The transverse acoustic phonon branch has an appreciable contribution to the diffuse scattering only near the zone center, along the azimuthal direction. These regions may thus be identified as being dominated by acoustic phonons. The regions of zero intensity near zone boundaries in Fig. S2a may be assumed to be dominated by optical phonons.

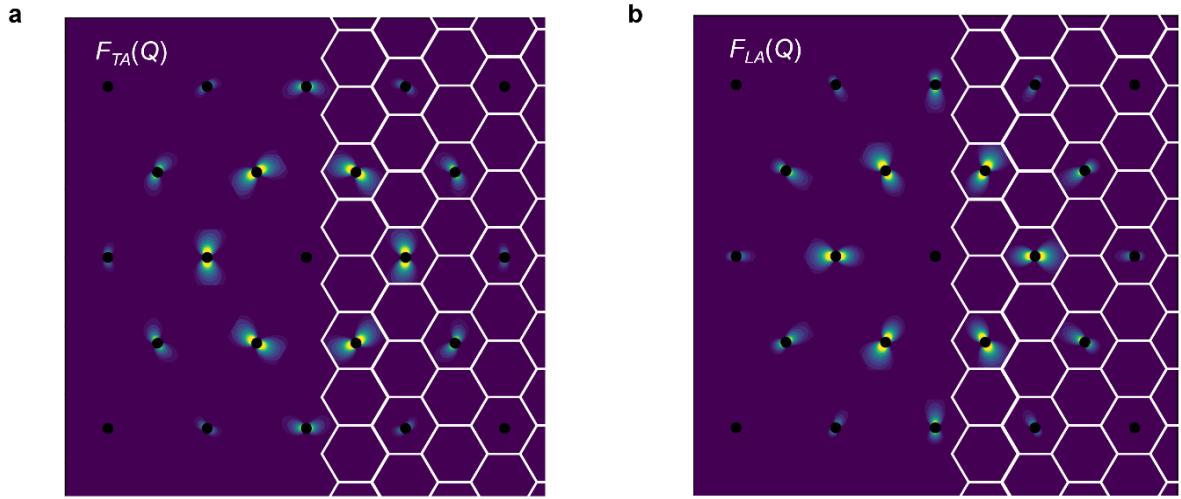

**Fig. S2 Simulation of structure factors of acoustic phonon branches. a,** Simulation of structure factor of the in-plane transverse acoustic (TA) phonon branch. **b,** Simulation of structure factor of the in-plane longitudinal acoustic (LA) phonon branch. Brillouin zones are overlayed in white as a guide to the eye. The structure factors are shown assuming a thermal population of the branches with a temperature of 300 K.

While it is possible that acoustic phonon branches may also contribute to diffuse scattering intensities at zone boundaries, it is useful to note that the classification of phonons into optical and acoustic branches is not meaningful away from the zone center. It is instead physically more appropriate to think of acoustic phonons near the zone center as being 'acoustic-like', and all phonons away from the zone center as being 'optical-like'. This is indeed the spirit of the approach used in the present work.

## S3. Non-thermal model of electron-phonon scattering

We interpret the UEDS data in terms of a non-thermal model of electron-phonon coupling, wherein hot, thermalized electrons couple at different rates to different phonon subsystems (here, optical and acoustic), each described by a distinct transient temperature. We use the method outlined in Section S2 to extract ultrafast pump-induced changes in optical and acoustic phonon populations. From Equation 1, $\Delta I_1 \propto \Delta n_i$, where $i$ is the dominant phonon branch (i. e. optical or acoustic) within an appropriate ROI. Furthermore, our measurements are carried out well above the Debye temperature (110 K, from reference[6]), allowing us to make use of the approximation $\hbar\omega_i \ll kT_i$, and thus $\Delta n_i \propto \Delta T_i$. Assuming that phonon populations are thermalized at long time delays, the above approach gives us ultrafast pump-induced temperatures of each phonon subsystem upto an overall normalization constant.

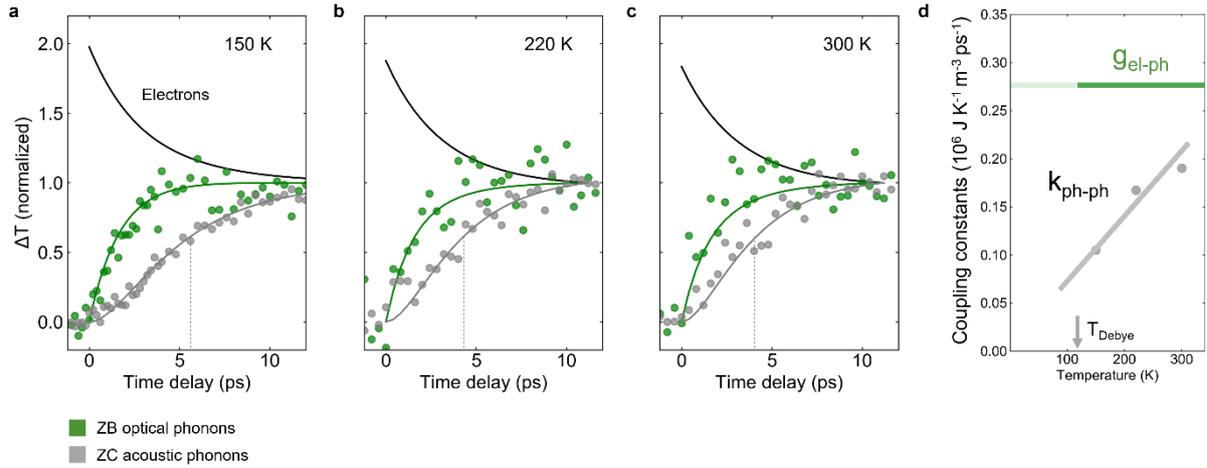

**Fig. S3 Non-thermal model of electron-phonon scattering.** Experimental zone-boundary (ZB) optical phonon and zone-center (ZC) acoustic phonon populations (circles) and fits to non-thermal model (solid lines), at three initial temperatures, **a,** 150 K, **b,** 220 K, and **c,** 300 K. Vertical dashed lines are used to illustrate the characteristic rise time of acoustic phonons. The electronic temperature is denoted by a solid black line. **d,** Temperature-dependent fit values for $g_{el-ph}$ and $k_{ph-ph}$. The grey line is a linear fit.

Next, we write down a set of coupled differential equations describing the non-thermal model, i. e. coupling of hot electrons to the two phonon subsystems. These are –

$$\dot{T}_e = -g_{el-ph}(T_e - T_{opt})/\gamma T_e, \quad (3)$$
$$\dot{T}_{opt} = \left(g_{el-ph}(T_e - T_{opt}) - k_{ph-ph}(T_{opt} - T_{ac})\right)/C_{opt}, \quad (4)$$
$$\dot{T}_{ac} = k_{ph-ph}(T_{opt} - T_{ac})/C_{ac}. \quad (5)$$

Here, $T_e$ is the electronic temperature, $T_{opt}$ is the optical phonon temperature, $T_{ac}$ is the acoustic phonon temperature, $g_{el-ph}$ is the electron-optical phonon coupling constant, $k_{ph-ph}$ is the phonon-phonon coupling constant, $\gamma$ is the coefficient of electronic specific heat (from reference), $C_{opt}$ is the specific heat of the optical phonon subsystem, and $C_{ac}$ is the specific heat of the acoustic phonon subsystem. According to the above equations, hot electrons are coupled only to the optical phonons. One may generally also consider an additional coupling to acoustic phonons, but as we discuss below, fitting of this model to our experimental results indicates that such a coupling is absent.

The specific heats of the optical and acoustic phonon subsystem are approximated analytically. The specific heat of phonon branch at frequency $\omega$ is given by $C = x \cdot e^x/(e^x - 1)^2$, where $x = \hbar\omega/k_B T$. Based on the phonon dispersions calculated using density functional theory[7] (also see Section S2 for a discussion on the distinction between optical and acoustic phonons), we assume an average $x = 0.5$ for optical phonons and $x = 0.05$ for acoustic phonons for simplicity. With 18 optical branches and 3 acoustic branches, we obtain $C_{opt} \sim 0.37 C_l$, and $C_{ac} \sim 0.63 C_l$, where $C_l$ is the lattice specific heat. We use $C_l = 3R$/mol, which is appropriate above the Debye temperature.

These approximations leave $g_{el-ph}$ and $k_{ph-ph}$ as the only unknown parameters, which we obtain by fitting to UEDS data obtained at 150 K, 220 K, and 300 K, shown in Fig. S3. The entire experimental dataset is fit to a single model, assuming a fixed, temperature-independent $g_{el-ph}$, and a temperature-dependent $k_{ph-ph}$. For the experiments at 150 K, the initial electronic temperature after pump excitation is approximately 1340 K, and the final temperature upon equilibration is around 750 K. The fit results are shown in Fig. S3 by solid lines.

The experimental acoustic phonon rate is dependent on the initial temperature, illustrated by the vertical dashed lines that denote the rise time. The optical phonon rate is independent of the initial temperature, within the experimental scatter. We find that the fit results are completely insensitive to the introduction of an additional coupling between electrons and acoustic phonons. This suggests that including such a coupling in the model is unphysical, i. e. the scattering rate between electrons and acoustic phonons is negligible.

## S4. Itinerant spin subsystem

The spin-polarized density of states projected onto Mn *3d*, Bi *5p*, and Te *4p* orbitals is shown in Fig. S4. The other orbitals near the Fermi level have a negligible contribution. ARPES and optical conductivity measurements have shown that the Fermi level in real crystals of MnBi$_2$Te$_4$ is ~0.2 eV above the conduction band edge, which we denote using a dashed line in Fig. S4. Examination of the density of states reveals that states at the Fermi level consist primarily of Bi *5p* and Te *4p* orbitals, which are spin-split in the both the nominal valence and conduction bands. The partial occupancy of the conduction band thus results in a finite itinerant magnetic moment. Integrating over the density of states, we find that the net itinerant magnetic moment is 0.022 $\mu_B$ due to Bi *5p* states and 0.011 $\mu_B$ due to Te *4p* states. The Mn bands do not contribute to the itinerant magnetic moment. We additionally note that while the net magnetic moment is in the conduction band, the optical excitation creates holes in the spin-split valence band, which will thus also contribute to the ultrafast spin dynamics. The valence band, like the conduction band, is also dominated by Bi *5p* and Te *4p* states.

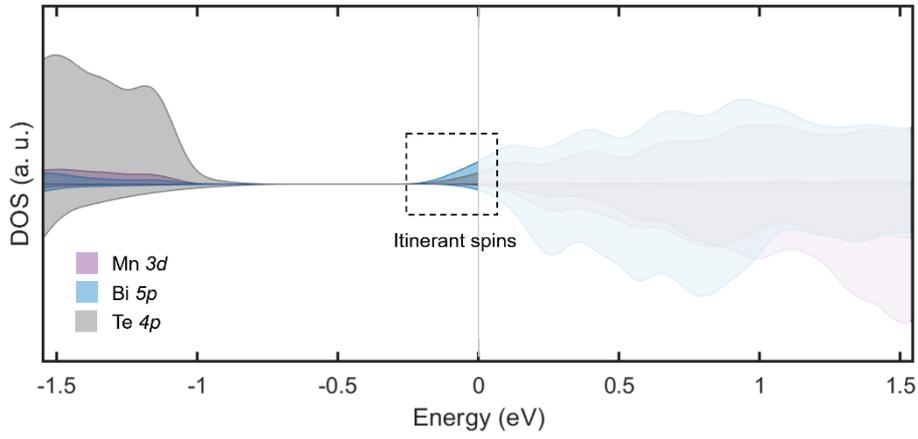

**Fig. S4 Itinerant spin bands.** The spin-polarized density of states projected onto different atomic orbitals near the Fermi level is plotted. The grey line denotes the experimental Fermi level. Occupied states are shaded. The dashed box shows the region of itinerant spins contributing to the net magnetic moment at equilibrium.

## S5. Transient pump-induced Kerr rotation and ellipticity

Ultrafast demagnetization measured via optical probes such as Kerr rotation may potentially suffer from artifacts due to dielectric bleaching, which can obscure true magnetic dynamics. An approach used in the literature to eliminate such artifacts is to simultaneously measure the pump-induced Kerr rotation $\Delta\theta$ as well as Kerr ellipticity $\Delta\eta$. In the case of true spin dynamics, $\Delta\theta$ and $\Delta\eta$ will track each other down to sub-picosecond timescales. We show in Fig. S5 that this is indeed the case in our measurements. In particular, within the first two picoseconds, where pump-induced carriers may potentially induce dielectric bleaching artifacts, $\Delta\theta$ and $\Delta\eta$ are identical.

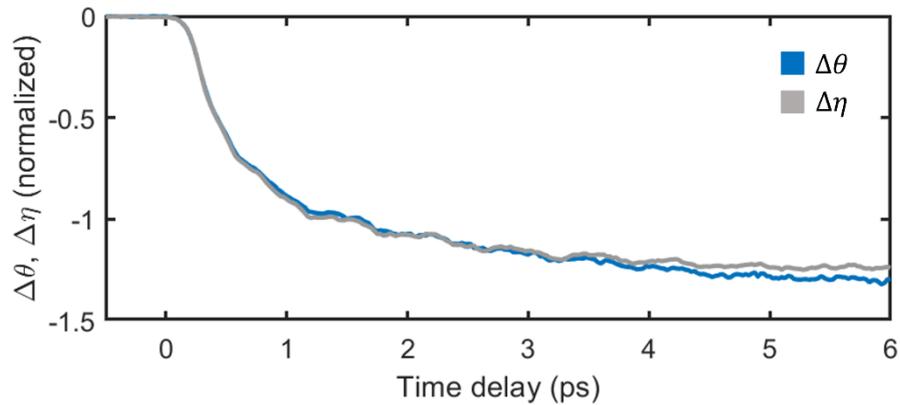

**Fig. S5 Transient Kerr rotation and ellipticity.** The pump-induced changes in Kerr rotation $\Delta\theta$ and ellipticity $\Delta\eta$ measured at 2 K, at a magnetic field of 3.8 T are plotted as a function of time delay.

## S6. Phenomenological model of electron-phonon spin-flip scattering

Electron-phonon spin-flip scattering of itinerant spins in MnBi$_2$Te$_4$ is modeled using a phenomenological model that considers individual electron-phonon scattering processes, following the approach in reference[8].

Hot electrons are assumed to remain internally in thermal equilibrium at all times, such that the electron occupation is described by a Fermi-Dirac distribution. Electron-phonon scattering is treated by the standard Fermi's golden rule approach. For simplicity, a single, dispersionless phonon branch of frequency $\omega$ is assumed, integrated over all scattering processes to obtain the phonon scattering rate. Finally, each electron-phonon scattering process is assumed to have a finite probability of flipping the spin of the electron, $a_{SF}$. The spins are treated as two level systems with an effective exchange splitting $\Delta_{ex}$.

The above physical processes can be described using the following coupled differential equations –

$$C_e \dot{T}_e = -C_p \dot{T}_p \quad (6)$$

$$\dot{N}_p = \frac{\pi}{\hbar} D_F^2 E_P \lambda_{ep}^2 \times \left( \coth\left(\frac{E_p}{2k_B T_e}\right) - \coth\left(\frac{E_p}{2k_B T_p}\right) \right) \quad (7)$$

$$\dot{S} = \frac{2\pi}{\hbar} \frac{D_F^2 D_p}{D_s} a_{SF} \lambda_{ep}^2 \times \begin{bmatrix} (1+N_p)\left(\frac{1}{2}-S\right)h(\Delta_{ex}+E_p,T_e) \\ -(1-N_p)\left(\frac{1}{2}+S\right)h(-\Delta_{ex}+E_p,T_e) \\ +N_p\left(\frac{1}{2}-S\right)h(\Delta_{ex}-E_p,T_e) \\ -N_p\left(\frac{1}{2}+S\right)h(\Delta_{ex}-E_p,T_e) \end{bmatrix} \quad (8)$$

Here, the $T_e$ and $T_p$ are the electronic and phonon temperatures, respectively. For simplicity, a single phonon temperature is considered, corresponding to optical phonons. $N_p$ is the phonon population, converted to a temperature $T_p$ using $N_p = \frac{1}{\exp\left(\frac{E_p}{k_B T_p}\right)-1}$. $C_e$ is the electronic specific heat, $C_p$ is the phonon specific heat, $D_F$ is the electronic density of states at the Fermi level, $D_p$ is the effective number of phonon branches, $E_p$ is an effective phonon energy, $D_s$ is the number of spins in each unit cell, $\lambda_{ep}$ is the microscopic electron-phonon coupling constant, $a_{SF}$ is the spin-flip probability, and $\Delta_{ex}$ is the effective exchange splitting of the itinerant spins. $S$ is the normalized average spin, and $\dot{S}$ is its time derivative. The function $h(x,T)$ is an integral over all allowed transitions from energy $E + x \to E$, integrated over $E$, in a Fermi-Dirac distribution at temperature $T_e$. This evaluates to $h(x,T) = \frac{x}{e^{\frac{E_p}{k_B T_e}}-1}$.

Several of the variables in Equations 6 – 8 are fixed from experiments and first-principles calculations. The electronic specific heat $C_e = \gamma T_e$, where $\gamma$ is $1.35 \times 10^3$ J m$^{-3}$ K$^{-1}$ from reference[6]. The phonon specific heat is approximated as a linear function in the temperature ranges considered here (10 K to 50 K), which is well below the Debye temperature, $C_p = \alpha T_p$, where $\alpha$ is $17.4 \times 10^3$ J m$^{-3}$ K$^{-1}$, derived from reference[6]. $D_F$ is approximated as 0.33 eV$^{-1}$ based on the experimental electronic specific heat of MnBi$_2$Te$_4$. Given that the temperatures considered here are well below the Debye temperature, $D_p$ is set to be 1, and $E_p$ to be 2.5 meV, which corresponds to a temperature of 30 K. We note that the choice of $D_p$ will only result in an overall scaling and will not affect the temporal dynamics. Finally, as described in the main text, MnBi$_2$Te$_4$ consists of localized as well as itinerant spin subsystems, located on different atoms. Since electron-phonon spin-flip processes directly influence only itinerant spins, we consider only the demagnetization of itinerant spins, which have a magnetic moment of 0.033 $\mu_B$ when the localized magnetic moment is fully saturated. This is in effect an assumption that the itinerant spins are isolated from the localized spins at ultrafast timescales. Finally, the magneto-optic Kerr rotation experiments are carried out at a magnetic field of 1 T, where the localized magnetic moment is ~0.1 of the saturated magnetic moment. We assume that the itinerant magnetic moment follows this, i. e. $D_s$ ~0.003 $\mu_B$ under our experimental conditions. We use the approach outlined in reference[9] to convert the experimental electron-optical phonon coupling constant from our UEDS measurements into a microscopic coupling constant, obtaining $\lambda_{ep}$ = 15 meV. The results of our UEDS measurements are thus directly used in modeling the ultrafast demagnetization.

This leaves $a_{SF}$, the spin-flip probability, and $\Delta_{ex}$, the effective exchange splitting as the only remaining variables. These variables are fit to the entire set of fluence-dependent time-resolved magneto-optic Kerr rotation data, shown in Fig. 3c. While we obtain physically reasonable values[9] of $a_{SF}$ = 0.06 and $\Delta_{ex}$ = 10 meV, we note that within our simplified model, these function as phenomenological parameters rather than an accurate estimation of physical quantities. This is primarily due to our simplifying assumption here that the itinerant spins are decoupled from localized Mn *3d* spins. Furthermore, the above model is nominally used to describe ferromagnets, whereas our experiments are in a spin-polarized paramagnetic state, under an external magnetic field.

The simplified model used here is mainly to illustrate that the observed ultrafast demagnetization of itinerant spins is consistent with electron-phonon spin-flip scattering. The role of the *p-d* exchange coupling and true antiferromagnetic dynamics are directly probed and discussed in detail in the final subsection of the Results section.

## S7. Resonant soft X-ray scattering

The soft X-ray reflectivity was measured as a function of energy to identify a resonance at the Mn $L_3$ edge. Measurements were carried out at 22.5° to satisfy the Bragg condition for the (0 0 0.5) magnetic peak, as described in the main text and Methods. The spectra measured at 10 K and 30 K in Fig. 4a show the sensitivity of the observed reflectivity signal to the antiferromagnetic (AFM) order.

We further confirm that the observed resonant reflectivity signal tracks the AFM order by measuring the reflectivity as a function of steady-state heating due to pump excitation. The results in Fig. S6 show that the reflectivity as a function of pump fluence exhibits the same characteristic shape as the temperature-dependent AFM order measured using neutron diffraction[10].

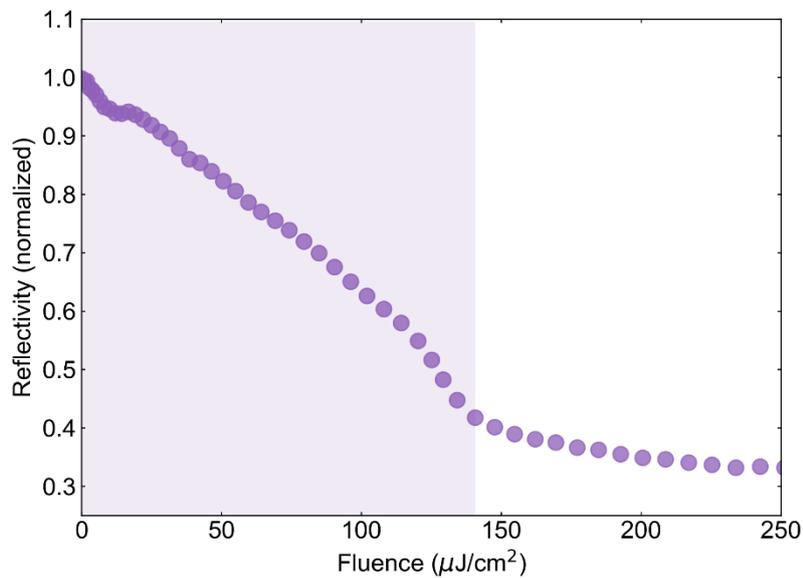

**Fig. S6 Resonant soft X-ray reflectivity dependence on steady state heating.** The reflectivity signal at 638 eV as a function of pump excitation fluence, at time delays before pump excitation. The sample temperature without laser pumping is 10 K. The purple shading highlights the region where the sample has finite antiferromagnetic order.

In the pump-probe data shown in Fig. 4b, the signal before pump excitation (i. e. time delay < 0) corresponds to the steady-state heating response shown in Fig. S6.

## S8. Three-temperature model of spin-lattice thermalization

The slow demagnetization dynamics in Fig. 4b are fit to a three-temperature model, consisting of electronic, lattice, and magnetic subsystems that thermalize with each other. The electronic and lattice temperatures are described as in the previous sections. The transient magnetic temperature is defined by mapping the time-dependent AFM order parameter $\langle S^2(t) \rangle$ to the equilibrium $\langle S^2 \rangle$ vs. temperature curve obtained from neutron scattering data in reference. This follows a power law dependence, given by $\langle S^2 \rangle \propto \left(1 - \frac{T}{T_N}\right)^n$, where $T_N = 24$ K, and $n = 0.7$.

The coupled differential equations describing such a thermalization can be written as –

$$\dot{T}_e = -g(T_e - T_l)/C_e, \qquad (9)$$
$$\dot{T}_l = \left(g(T_e - T_l) - g_{sl}(T_l - T_m)\right)/C_l, \qquad (10)$$
$$\dot{T}_m = g_{sl}(T_l - T_m)/C_m, \qquad (11)$$

where $T_e$, $T_l$, and $T_m$ are the electronic, lattice, and magnetic temperatures respectively, $g$ is the electron-phonon coupling constant, $g_{sl}$ is the spin-lattice coupling constant, and $C_e$, $C_l$, and $C_m$ are the electronic, lattice and magnetic specific heats respectively.

The electron-phonon coupling constant $g_{el-ph}$ is set to the value obtained from our UEDS measurements. $C_e$ and $C_l$ are estimated as described in the Section S6. $C_m$ is calculated from specific heat measurements in reference as ~1×10³ J m⁻³ K⁻¹. The set of fluence-dependent time-resolved soft X-ray reflectivity data in the low fluence limit is fit to a single $g_{sl}$.

## S9. Temperature-dependent trMOKE measurements

The scenario of two distinct demagnetization processes with disparate timescales described in Fig. 4 of the main text is further supported by temperature-dependent trMOKE measurements. The results, plotted in Fig. S7a, show two clear regimes in the time domain. In the electron-phonon scattering regime (labeled 'el-ph'), occurring at timescales <15 ps, ultrafast demagnetization is observed with a characteristic shoulder that is consistent with electron-optical phonon scattering, and independent of the initial temperature. In the spin-lattice thermalization regime (labeled 'thermalization'), the transient magnetization dynamics exhibit a dramatic temperature dependence. At temperatures below $T_N = 24$ K, the transient magnetization increases with time delay, whereas above $T_N$, the transient magnetization decreases, occurring over timescales of hundreds of picoseconds, consistent with the RSXS measurements. The behavior in the spin-lattice thermalization regime is explained upon examination of the equilibrium magnetization ($M$) vs. temperature ($T$) curve, plotted in Fig. S7b. Under an out-of-plane magnetic field of 1 T, $M$ exhibits a divergence at $T_N$, characteristic of the magnetic susceptibility of antiferromagnets. The magnetic dynamics in the spin-lattice thermalization regime in Fig. S7a follow the equilibrium $M$ vs. $T$ curve, as illustrated by the arrows in Fig. S7b. The stark contrast in temperature- and fluence-dependent magnetic dynamics in the two regimes described above unambiguously confirm the distinct demagnetization processes for itinerant and localized spins.

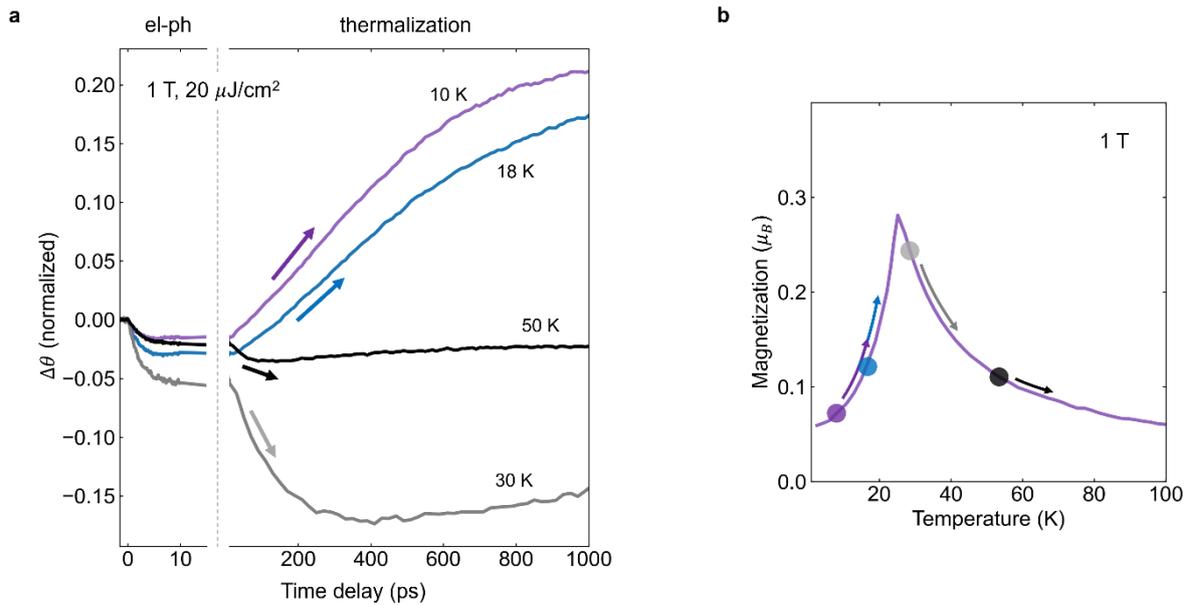

**Fig. S7 Temperature-dependent trMOKE measurements. a,** Temperature-dependent time-resolved magneto-optic Kerr rotation, with the two regimes of demagnetization dynamics labeled on top. **b,** Static magnetization ($M$) vs. temperature ($T$) measurement (solid purple line). Circles (color-coded to the plots in panel (a)) are used to denote the equilibrium $M$ and $T$ for each of the plots in panel (a). The arrows in both (a) and (b) denote the evolution of the time-dependent magnetization in the spin-lattice thermalization regime highlighted in panel (a).

## S10. Landau-Lifshitz-Gilbert simulations

The Landau-Lifshitz-Gilbert (LLG) simulations were carried out for the two spin subsystems assuming an elevated electronic temperature at $t = 0$, followed by a subsequent thermalization between the electrons and lattice. The initial electronic temperature was calculated for the considered pump fluences using the experimental specific heat as in Section S5, the electron-lattice thermalization timescale was set by our UEDS results in Section S3, and the final lattice temperature calculated using the experimental specific heat as in Section S5.

We use previously reported experimental results[10,11] to fix the values of various parameters in our Landau-Lifshitz-Gilbert (LLG) simulations. We set $J_{d,OOP}$ = -0.022 meV, $K_d$ = 0.05 meV, and $S_d$ = 2.5, and estimate $S_p$ = 0.0025 using density functional theory calculations (see Section S4).

The value of $J_{d,IP}$ = 0.16 meV is set by reproducing the static antiferromagnetic order as a function of temperature, as shown in Fig. S8a. The simulation reproduces the characteristic antiferromagnetic order parameter behavior, with the correct ordering temperature of $T_N$ = 24 K.

We determine $\alpha_d$ by setting $J_{pd} = 0$ and matching the spin-lattice thermalization dynamics at low pump excitation fluences, where electron-phonon spin-flip scattering is negligible. We find a good agreement with the experimental results for $\alpha_d$ = 0.003, as shown in Fig. S8b. In particular, the simulated and experimental dynamics clearly correspond to a single exponential decay, consistent with spin-lattice thermalization.

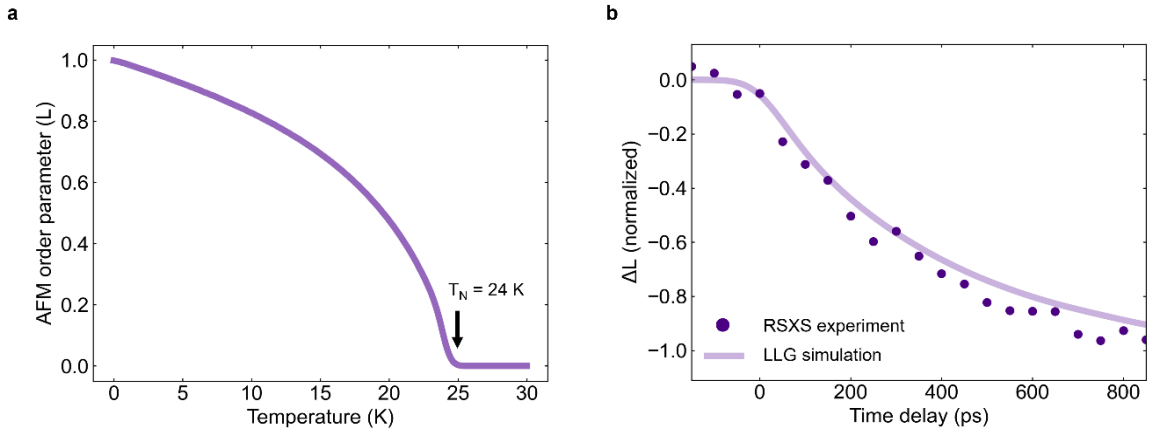

**Fig. S8 Landau-Lifshitz-Gilbert simulations. a,** The static antiferromagnetic order ($L$) as a function of temperature using Landau-Lifshitz-Gilbert (LLG) simulations. The experimental ordering temperature $T_N$ is marked with a black arrow. **b,** Pump-induced change in the antiferromagnetic order parameter ($\Delta L$) from LLG simulations (solid line) and resonant soft X-ray (RSXR) scattering experiments.

Finally, we determine $\alpha_p$ = 0.5 and $J_{pd}$ in the range of 10-50 meV by comparing the simulation results with the experimental results at intermediate fluences as described in the main text.

1. Padmanabhan, H. *et al.* Interlayer magnetophononic coupling in MnBi2Te4. (2021).
2. Xu, R. & Chiang, T. C. Determination of phonon dispersion relations by X-ray thermal diffuse scattering. *Zeitschrift fur Krist.* **220**, 1009–1016 (2005).
3. René De Cotret, L. P. *et al.* Time- and momentum-resolved phonon population dynamics with ultrafast electron diffuse scattering. *Phys. Rev. B* **100**, (2019).
4. Seiler, H. *et al.* Accessing the anisotropic non-thermal phonon populations in black phosphorus. *arXiv* **1**, (2020).
5. Chase, T. *et al.* Ultrafast electron diffraction from non-equilibrium phonons in femtosecond laser heated Au films. *Appl. Phys. Lett.* **108**, (2016).
6. Yan, J. Q. *et al.* A-type antiferromagnetic order in MnBi4Te7 and MnBi6Te10 single crystals. *Phys. Rev. Mater.* **4**, 1–14 (2020).
7. Li, J. *et al.* Intrinsic magnetic topological insulators in van der Waals layered MnBi 2 Te 4 -family materials. *Sci. Adv.* **5**, eaaw5685 (2019).
8. Longa, D. *Laser-induced magnetization dynamics : an ultrafast journey among spins and light pulses*. (2008). doi:10.6100/IR635203
9. Koopmans, B. *et al.* Explaining the paradoxical diversity of ultrafast laser-induced demagnetization. *Nat. Mater.* **9**, 259–265 (2010).
10. Yan, J. Q. *et al.* Crystal growth and magnetic structure of MnBi2Te4. *Phys. Rev. Mater.* **3**, 1–8 (2019).
11. Li, B. *et al.* Competing Magnetic Interactions in the Antiferromagnetic Topological Insulator MnBi2Te4. *Phys. Rev. Lett.* **124**, 1–6 (2020).